\documentclass[11pt]{article}

\usepackage{stolenstyle}

\usepackage{amsmath,epsf,amssymb,latexsym,amsthm,setspace,bbm,array,pifont,multirow}

\usepackage[matrix,arrow,frame,import,curve,color]{xy}

\DeclareFontFamily{U}{rsf}{}
\DeclareFontShape{U}{rsf}{m}{n}{
  <5> <6> rsfs5 <7> <8> <9> rsfs7 <10-> rsfs10}{}
\DeclareMathAlphabet\Scr{U}{rsf}{m}{n}

\def\cDb{{\overline{\cD}}}

\def\C{{\mathbb C}}

\def\P{{\mathbb P}}
\def\R{{\mathbb R}}
\def\Z{{\mathbb Z}}

\def\Aut{\operatorname{Aut}}

\def\diag{\operatorname{diag}}

\def\rank{\operatorname{rank}}

\def\GL{\operatorname{GL}}

\def\GU{\operatorname{U{}}}

\def\la{\langle}
\def\ra{\rangle}

\def\cD{{\cal D}}
\def\cE{{\cal E}}

\def\cL{{\cal L}}

\def\cO{{\cal O}}

\def\ep{{\epsilon}}

\newcommand\alphah{\widehat{\alpha}}
\newcommand\betah{\widehat{\beta}}

\newcommand\kappah{\widehat{\kappa}}

\newcommand\muh{\widehat{\mu}}

\newcommand\rhoh{\widehat{\rho}}

\newcommand\ah{\widehat{a}}
\newcommand\bh{\widehat{b}}

\newcommand\jh{\widehat{\jmath}}
\newcommand\kh{\widehat{k}}

\newcommand\Dh{\widehat{D}}

\newcommand\Qh{\widehat{Q}}

\newcommand\Jt{\widetilde{J}}

\newcommand\Mt{\widetilde{M}}

\makeatletter
\newsavebox{\@brx}
\newcommand{\llangle}[1][]{\savebox{\@brx}{\(\m@th{#1\langle}\)}%
  \mathopen{\copy\@brx\kern-0.5\wd\@brx\usebox{\@brx}}}
\newcommand{\rrangle}[1][]{\savebox{\@brx}{\(\m@th{#1\rangle}\)}%
  \mathclose{\copy\@brx\kern-0.5\wd\@brx\usebox{\@brx}}}
\makeatother

\def\bX{{{\boldsymbol{X}}}}
\def\bsigma{{{\boldsymbol{\sigma}}}}
\def\bSigma{{{\boldsymbol{\Sigma}}}}
\def\bGamma{{{\boldsymbol{\Gamma}}}}
\def\bdelta{{{\boldsymbol{\delta}}}}
\def\bgamma{{{\boldsymbol{\gamma}}}}

\def\be{{{{\boldsymbol{e}}}}}

\def\e{{{\mathfrak{e}}}}

\def\preprint{IPMU18-0108}

\title{Testing the (0,2) mirror map}
\author{Marco Bertolini}
\affiliation{Kavli Institute for the Physics and Mathematics of the Universe (WPI), \\
The University of Tokyo Institutes for Advanced Study,\\ 
The University of Tokyo, Kashiwa, Chiba 277-8583, Japan}
\emailAdd{marco.bertolini@ipmu.jp}

\abstract{We test a proposed mirror map at the level of correlators for linear models describing the (0,2) moduli space of superconformal field theories
with a (2,2) locus associated to Calabi-Yau hypersurfaces in toric varieties.
We verify in non-trivial examples that the correlators are exchanged by the mirror map and 
we derive a correspondence between the observables of the A/2- and B/2-twisted theories. We also comment on the 
global structure of the (0,2) moduli space and present a simple non-renormalization argument for a large class of B/2 model
subfamilies.}

\begin{document}

\maketitle

\section{Introduction}
\label{s:intro}
 
In this work, we further the study of (0,2) mirror symmetry \cite{Melnikov:2012hk} for deformations of (2,2) theories.
That such a (0,2) mirror map exists -- as opposed to (0,2) models without a (2,2) locus -- is guaranteed by (2,2) mirror symmetry 
\cite{Greene:1990ud,Candelas:1990rm}:
two isomorphic (2,2) superconformal field theories (SCFTs) have, by definition, isomorphic spaces of deformations.
In what follows, we restrict our attention to such deformations which preserve at least (0,2) supersymmetry.

A particularly prominent role in mirror symmetry investigations has been played by 
the gauged linear sigma model (GLSM) \cite{Witten:1993yc}. This is a two-dimensional supersymmetric gauge theory
which, for a suitable choice of its parameters, flows in the IR to a non-linear sigma model (NLSM) with
target space a complete intersection Calabi-Yau (CICY) in a toric variety. 
Some of the parameters of the IR conformal field theory have natural representatives
in the UV linear model. These are encoded in two types of superpotentials: a chiral superpotential
encodes the complex structure parameters and a twisted-chiral superpotential
encodes the K\"ahler parameters. These generically describe only a subset of 
the full space of deformations of the CFT, as they do not include non-polynomial
complex structure deformations -- turning these on would prevent the CICY to
be embedded in the toric variety --  or non-toric K\"ahler deformations -- 
these are associated to divisors on the ambient toric variety 
that become reducible when restricted to the CICY --.

A remarkable property of (2,2) mirror symmetry is that it respects this toric structure on both sides of the mirror, that is, 
polynomial complex structure deformations are mapped to toric K\"ahler deformations of the mirror and vice versa. 
This is encoded in the monomial-divisor mirror map (MDMM) \cite{Aspinwall:1993rj}.

(2,2) SCFTs admit deformations which preserve only (0,2) supersymmetry. In a large radius phase 
these are interpreted as deformations of the tangent bundle over the CICY.
For a subset of these there exist corresponding holomorphic deformations of the linear model Lagrangian,
and we refer to these as {\it monad} deformations. These, together with polynomial complex structure and K\"ahler deformations,
form the toric moduli space of the SCFT.\footnote{The concept of a toric moduli space is not universal, as there are
different GLSM realizations of the same SCFT realizing torically different subsets of the moduli space of the conformal theory \cite{Distler:1995bc,Distler:1996tj}.} 
It therefore comes naturally to ask whether mirror symmetry respects the toric structure of this enlarged (0,2) moduli space,
that is, if monad deformations of the tangent bundle are exchanged by the mirror map.
This turns out not to be the case. In fact, a careful analysis of the GLSM parameters modulo field 
redefinitions shows that in general even the dimensions of the toric moduli space of the theory and its mirror
do not match \cite{Kreuzer:2010ph}. 

However, there exist subfamilies of such theories which appear to be exchanged by mirror symmetry.
A proposal for an extension of the MDMM to a class of (0,2) deformations of (2,2) linear models 
for CY hypersurfaces was presented in \cite{Melnikov:2010sa}. 
When this sublocus coincides with the full parameter space of the theory, the model is referred to
as {\it reflexively plain} \cite{Kreuzer:2010ph}. 
It has been shown that the map satisfies non-trivial tests. In fact, the map exchanges the dimensions of the (subloci of the) moduli spaces of the mirror theories, 
as well as it exchanges the principal component of the singular locus of the A/2-twisted theory with the principal component of the singular
locus of the B/2-twisted mirror theory.

In this work, we will further test the mirror map, and we will show that it leads to an isomorphism at the level of the correlators. 
That is, the map exchanges correlation functions in the A/2-twisted theory \cite{McOrist:2008ji} with correlation functions 
in the B/2-twisted mirror theory. As a consequence, we derive a correspondence between the natural observables
of the twisted theories on both sides of the mirror. Unfortunately, we still lack the technical tools to carry out such analysis 
for a generic model. While recently there has been progress on the A model \cite{Closset:2015ohf,Closset:2017vvl} 
and A/2 model \cite{Closset:2015rna} side, techniques to evaluate B model correlators mostly rely on the existence of 
limiting points where the theory can be solved. For this reason, we focus on theories which admit a Landau-Ginzburg orbifold (LGO) phase.
As it has been argued that the corresponding B/2 model does not receive instanton corrections \cite{McOrist:2008ji}, 
the relevant B/2-twisted correlators are therefore independent of the phase structure of the theory,
and we can make use of the LGO description to solve the model. 
As a consequence, we will be able to argue that the B/2 model of any theory to which the mirror map applies
does not receive contributions form worldsheet instantons, regardless of the existence of a LGO phase.

The rest of this paper is organized as follows. In section \ref{s:setup}, we review just enough basic notions of toric geometry and of linear models in order to
present the (0,2) mirror map of \cite{Melnikov:2010sa}. In section \ref{s:quinticorb}, we solve a reflexively plain example, while in section \ref{s:octic}
we tackle a more general model, where the mirror map acts on appropriate subfamilies of the moduli space. We conclude with some 
comments on the structure of the (0,2) moduli space as well as some open questions.

\acknowledgments It is a pleasure to thank I.~Melnikov and R.~Plesser for helpful discussions and comments on the manuscript. 
This work was supported by World Premier International Research Center Initiative (WPI Initiative), MEXT, Japan.

\section{The (0,2) mirror symmetry set-up}
\label{s:setup}

In this section we review the GLSM construction of mirror pairs for Calabi-Yau hypersurfaces in toric varieties, introducing along the
way some relevant toric geometry notions. As this material is quite standard, our discussion will not try to be exhaustive, and for more details 
the reader can for instance refer to \cite{Cox:2000vi}.

\subsection{Polytopes and hypersurfaces}

We start with a $d$-dimensional lattice polytope $\Delta\in M_{\R}\simeq \R^d$, and 
we define the dual polytope 
\begin{align}
\Delta^\circ \equiv \left\{ \rho\in N_{\R} | \la m , \rho \ra \geq-1, \ \forall m\in\Delta \right\}~,
\end{align}
where $N_{\R}\simeq (M_{\R})^\vee$ and the pairing product $\la \cdot , \cdot \ra: M_{\R}\times N_{\R}\rightarrow\R$ descends from the standard product in $\R^d$.
A lattice polytope $\Delta$ is said to be reflexive if and only if its dual $\Delta^\circ$ is also a lattice polytope, and it follows
that both $\Delta$ and $\Delta^\circ$ have a unique interior point which we assume, without loss
of generality, to be the origin.

Let $\Delta$ be a reflexive polytope. Then $\Delta$ can be interpreted as the Newton polytope for a hypersurface 
$\{P=0\}\in V$, where $V$ is the toric variety whose fan $\Sigma_V\in N_{\R}$ is obtained by taking cones over the faces of $\Delta^\circ$. 
To make this more explicit, let $\rho_1,\dots,\rho_n$ be the non-zero points in $\Delta^\circ$, which correspond to integral generators 
of the one-dimensional cones of $\Delta^\circ$.\footnote{This corresponds to a refinement of the original fan known as a maximal projective subdivision,
and note that it is not unique. The corresponding toric variety is obtained by a toric crepant resolution of singularities, and although it
will not be in general smooth, its singularities are sufficiently mild \cite{Cox:2000vi}.}
For each of these we introduce a coordinate $X_\rho\in\C^n$, and we present the toric variety $V$ as the holomorphic quotient
\begin{align}
V=\frac{\C^n-Z(F)}{G}~,
\end{align} 
where $G=(\C^\ast)^r\times H$, $r=n-d$ and $H$ is a finite Abelian group. 
$Z(F)\in\C^n$ is the subvariety associated to the irrelevant ideal (or Cox ideal) in the homogeneous 
coordinate ring $\C[X_{\rho_1},\dots,X_{\rho_n}]$, and in particular it depends on the specific triangulation $\Sigma_V$. 
The $(\C^\ast)^r$ action on the coordinates $X_\rho$ is determined in terms of a basis for the kernel of the pairing matrix\footnote{In order to 
simplify notation, we will often assume a specific ordering of the points in $\Delta$ and write $\rho$ instead of $\rho_i$.}
 $\la m,\rho\ra$
between the non-zero elements of $\Delta$ and $\Delta^\circ$. The pairing matrix has rank $d$
by construction, and the quotient action is given by
\begin{align}
\label{eq:quotaction}
X_\rho &\rightarrow \prod_a (\lambda_a)^{Q^a_\rho} X_\rho~,		&\lambda_a\in(\C^\ast)^r~,
\end{align}
where the charges $Q^a_\rho$ span an integral basis for the kernel of $\la m,\rho\ra$. 

The polynomial defining the CY hypersurface is given in terms of the homogeneous coordinates by
\begin{align}
\label{eq:hypeq}
P(X) = \sum_{m\in\Delta} \alpha_m \prod_{\rho} X_\rho^{\la m,\rho\ra+1}~.
\end{align}
In particular, under the action \eqref{eq:quotaction} each monomial in $P$ transforms according to
\begin{align}
\prod_\rho X_\rho^{\la m,\rho\ra+1} \rightarrow \left( \prod_a \lambda_a^{\sum_\rho Q^a_\rho}\right) \prod_\rho X_\rho^{\la m,\rho\ra+1}~,
\end{align}
and therefore \eqref{eq:hypeq} transforms with charge $\sum_\rho Q^a_\rho$ and the Calabi-Yau hypersurface $M=\{P=0\}\subset V$ is well-defined.  

The fact that $\Delta$ is reflexive is equivalent to the fact that $\Delta^\circ$ is reflexive as well. Therefore, applying the procedure we summarized above
while exchanging the roles of $\Delta$ and $\Delta^\circ$, we obtain a Calabi-Yau hypersurface $M^\circ=\{P^\circ=0\}\subset V^\circ$. In particular, 
let $m$ label the non-zero points in $\Delta$ and $n^\circ=|m|$, and let us introduce a homogeneous coordinate $Y_m\in\C^{n^\circ}$ for each of these.
Then we have
\begin{align}
V^\circ=\frac{\C^{n^\circ}-Z(F^\circ)}{G^\circ}~,
\end{align} 
where $(\C^\ast){}^{r^\circ}\times H^\circ$, $r^\circ=n^\circ-d$ and $F^\circ$ is the irrelevant ideal in the homogeneous coordinate ring $\C[Y_{m_1},\dots,Y_{m_{n^\circ}}]$. 
The $(\C^\ast){}^{r^\circ}$ action on these is given by
\begin{align}
\label{eq:quotactionmirr}
Y_m &\rightarrow \prod_{\ah} (\lambda_{\ah})^{\Qh^{\ah}_m} Y_m~,		&\lambda_{\ah}\in(\C^\ast)^{r^\circ}~,
\end{align}
where $\Qh^{\ah}_m$ is an integral basis for the kernel of the pairing matrix $\la \rho,m \ra$ for non-zero $\rho\in\Delta^\circ$ and $m\in\Delta$, 
which is simply the transpose of the pairing matrix $\la m,\rho\ra$ above. Thus, $\Qh^{\ah}_m$ parametrize an integral basis for the cokernel of $\la m,\rho\ra$.
Similarly, one can show that
\begin{align}
\label{eq:hypeqmirr}
P^\circ(Y) = \sum_{\rho\in\Delta^\circ} \alphah_\rho \prod_m Y_m^{\la \rho,m\ra+1}
\end{align}
transforms under \eqref{eq:quotactionmirr} with charge $\sum_m \Qh^{\ah}_m$, and the 
Calabi-Yau hypersurface $M^\circ=\{P^\circ=0\}\subset V^\circ$ is well-defined as well.

In the context of a heterotic string theory background, a (0,2) NLSM is constructed by specifying a target space $M$,
which we assume to be of the form described above,
together with a holomorphic vector bundle $\cE\rightarrow M$. 
When $\cE=T_{M}$ the theory possesses (2,2) supersymmetry 
and the pairs $T_M\rightarrow M$ and $T_{M^\circ}\rightarrow M^\circ$ form a Batyrev mirror pair \cite{Batyrev:1994hm}
and constitute the starting point for our analysis.
These theories admit deformations that preserve only (0,2) supersymmetry. In the geometric phase, these correspond
to a bundle $\cE$ obtained as a deformation of the tangent bundle.
As mentioned above, for a given UV GLSM realization of the low-energy SCFT, 
the associated (0,2) linear model moduli space includes, 
in addition to the toric K\"ahler and polynomial complex structure deformations,
a subset of the bundle deformations.
This is the moduli space to which the (0,2) mirror map applies and to which we turn in the next section.

\subsection{The (0,2) linear model}

We now turn to the physical theory of interest, namely the GLSM. 
This is a two dimensional supersymmetric gauge theory, whose gauge group we assume to be Abelian in this work.
For a suitable choice of values for the UV parameters, the theory flows in the IR to a NLSM describing the geometric structure 
we introduced above.

Such theory is constructed with $n+1$ (0,2) bosonic chiral superfields $X_0,X_\rho$ and the same number of
(0,2) Fermi multiplets $\Gamma^0,\Gamma^\rho$. These have gauge charges $Q^a_0\equiv-\sum_\rho Q^a_\rho, Q^a_\rho$
and they are coupled to a collection of $r$ gauge-neutral chiral supermultiplets $\Sigma_a$. The chirality conditions
for the Fermi fields read
\begin{align}
\label{eq:Gammachircond}
\cDb \Gamma^0 &= \sum_a \Sigma_a X_0 E_0(X)~,		&\cDb \Gamma^\rho &= \sum_a \Sigma_a E_\rho(X)~,
\end{align}
where $E_0(X),E_\rho(X)$ are polynomials in the superfields $X_\rho$, and $\cDb$ is a gauge covariant (super)derivative.\footnote{We follow
the conventions of \cite{Bertolini:2014dna} for (0,2) linear models, to which we refer for more details.}

The action for the theory is determined by the kinetic terms for the gauge fields $V_a$, for the $\Sigma_a$ fields and for the various matter fields, 
as well as by superpotential interactions 
\begin{align}
\cL_W &= \int d\theta \left[ \sum_\rho X_0 \Gamma^\rho J_\rho(X) + \Gamma^0 H(X) \right] + \text{h.c.}~, \nonumber\\
\cL_{\text{F.I.}} &=-\frac14 \int d\theta \tau^a \Upsilon_a + \text{h.c.}~,
\end{align}
where $\Upsilon_a$ are the twisted-chiral gauge-invariant field strengths and $\tau^a=ir^a+\theta^a/2\pi$ are the complexified F.I.~parameters.
The parameters $q_a=e^{2\pi i\tau^a}$ correspond to the algebraic coordinates on the complexified K\"ahler moduli space of $V$
\cite{Morrison:1994fr,Morrison:1995yh}, while
$P(X)$ is determined by \eqref{eq:hypeq}. The remaining functions $J_\rho(X)$ are generic polynomials in the fields $X_{\rho'}$
with gauge charges $-Q^a_0-Q^a_\rho$. In particular, $J_\rho(X)$ are allowed to contain all the monomials that 
are allowed to appear in $\partial_\rho P$, that is
\begin{align}
\label{eq:XJsexpre}
X_\rho J_\rho = \sum_{m\in\Delta} j_{m\rho} \prod_{\rho'} X_{\rho'}^{\la m,\rho' \ra+1}~.
\end{align}
Note that in this expression $\rho'\neq0$ while $m$ sums over all points in $\Delta$, including the origin.
The condition that the RHS of \eqref{eq:XJsexpre} is divisible by $X_\rho$ implies that $j_{m\rho}=0$ whenever $\la m,\rho\ra=-1$.
The (2,2) locus is described by $j_{m\rho}=\alpha_m (\la m,\rho\ra+1)$.

The map specified in \cite{Melnikov:2010sa} is restricted to a subset of the GLSM moduli space. 
This is realized by constraining the polynomials $E(X)$ in \eqref{eq:Gammachircond}.
Following the same notation of \cite{Melnikov:2010sa}, we restrict our attention to the following form of the chirality conditions
for the left-moving Fermi fields
\begin{align}
\label{eq:Esdiagform}
\cDb \Gamma^0 &= X_0 \bSigma \cdot \bdelta~,		&\cDb \Gamma^\rho &= X_\rho \bSigma \cdot \be^\rho~,
\end{align}
where $\bSigma$ is a vector with components the fields $\Sigma_a$, while
$\bdelta$ and $\be^\rho$ are vectors of parameters of dimension $r$.

The GLSM action is (0,2) supersymmetric if and only if
\begin{align}
\bdelta H(Z) + \sum_\rho \be^\rho X_\rho J_\rho=0~,
\end{align}
which using \eqref{eq:XJsexpre} can be recasted as
\begin{align}
\sum_{m\in\Delta} \left[ \alpha_{m} \bdelta+ \sum_\rho \be^\rho j_{m\rho}\right]  \prod_{\rho'}X_{\rho'}^{\la m,\rho' \ra+1}=0~,
\end{align}
which in turn implies the relations
\begin{align}
\label{eq:02SUSYconstr}
\alpha_{m} \bdelta+ \sum_\rho \be^\rho j_{m\rho}=0~,	\qquad\qquad \forall m\in\Delta~.
\end{align}
If we assume a triangulation $\Sigma_V$ corresponding to a large radius phase of the linear model, 
the space of classical vacua is the Calabi-Yau hypersurface $M=\{ X\in V | P(X)=0 \}\subset V$, where
by a slight abuse of notation we denote by $X_\rho$ the lowest components of the corresponding bosonic superfields. 
The left-moving fermions of the theory, which appear as the lowest components in the Fermi supermultiplets $\Gamma^\rho$,
couple to the holomorphic bundle $\cE$ defined by the cohomology of the short sequence
\begin{align}
\label{eq:bundlecohom}
\xymatrix@R=0mm@C=10mm{
0 \ar[r]	&\cO^{\oplus r} \ar[r]^-{\be^\rho X_\rho} &\oplus_\rho \cO(Q^a_\rho)\ar[r]^-{J_\rho} & \cO(-Q^a_0) \ar[r]	&0
}
\end{align}
restricted to $M$. For the class of models we are considering, the holomorphic vector bundle $\cE \rightarrow M$,
which we assume nonsingular, describes a subset of deformations of the tangent bundle $T_M$.

Of course, the same construction obtained by exchanging the data as described in the previous section 
yields the mirror GLSM theory, which flows in the IR to a NLSM for the geometry $\cE^\circ \rightarrow M^\circ$.

\subsection{The (0,2) mirror map}

Having reviewed the class of models which will be the subject of our study, we are ready to 
state the (0,2) mirror map proposal of \cite{Melnikov:2010sa}. 
Upon restricting ourselves to the subset of the parameter space identified by \eqref{eq:Esdiagform}, it is possible to 
find a parametrization of the linear model which is invariant under the field redefinitions 
corresponding to the $(\C^\ast)^d$ subgroup of $\Aut V$, whose action simply rescales 
the coordinates of $V$. We denote these as ``toric" field redefinitions.
Of course, for a generic model the automorphism group of $V$ is
larger, and the description we review below is clearly an overparametrization
of the parameter space. However, this redundancy is mirror symmetric \cite{Melnikov:2010sa} and,
as in the (2,2) case, the (0,2) mirror map naturally extends to this redundant description. 
Hence, this fact does not lead to any difficulties in our task of computing the correlators in both theories.
In fact, we will see that allowing for some additional redundancy will make the result look simpler. 

The generalizations of invariant ``complex structure" and ``K\"ahler" coordinates, respectively, are given by
\begin{align}
\label{eq:cxpkaehdef}
\kappah_{\ah} &\equiv \prod_{m\neq0} \left( \frac{\alpha_m}{\alpha_0}\right)^{\Qh^{\ah}_m}~,	&\kappa_a&\equiv q_a \prod_\rho \left( \frac{j_{0\rho}}{\alpha_0}\right)^{Q^a_\rho}~.
\end{align}
Notice that $\kappah_{\ah}$ coincide with the complex structure coordinates of the (2,2) theory, while on the (2,2) locus, $\kappa_a=q_a$.
For the bundle data, a toric field redefinition invariant quantity is given by
\begin{align}
b_{m\rho}&\equiv \frac{\alpha_0 j_{m\rho}}{\alpha_m j_{0\rho}}-1~,		 &&\text{for }m\neq0~,
\end{align}
subject to the condition that $b_{m\rho}=-1$ whenever $\la m,\rho \ra=-1$.
Assuming that $\alpha_m\neq0$, it follows from the (0,2) SUSY constraints \eqref{eq:02SUSYconstr} that
\begin{align}
\label{eq:gamdeltconstr}
\bdelta &= -\sum_\rho \bgamma^\rho~,		&\sum_\rho b_{m\rho}\bgamma^\rho &=0~,		
\end{align}
where we defined
\begin{align}
\bgamma^\rho \equiv \frac{j_{0\rho}}{\alpha_0}\be^\rho~.
\end{align}
The vectors $\bdelta$ and $\bgamma^\rho$, which therefore span the kernel of $b_{m\rho}$, are determined by \eqref{eq:gamdeltconstr}
only up to a $\GL(r,\C)$ transformation
corresponding to the field redefinitions for the $r$ $\Sigma_a$ multiplets. In fact, the theory is singular both if
$\rank b_{m\rho}>d$, which corresponds to a vanishing $\bgamma^\rho$ and a free $\Sigma_a$ multiplet,
as well as if $\rank b_{m\rho}<d$, in which case the B/2-twisted theory develops a singularity. Thus we restrict 
our attention to $b_{m\rho}$ having exactly rank $d$. This means that $\dim \ker b_{m\rho}=r$ and that the $\bgamma^\rho$
are completely determined up to the $\GL(r,\C)$ field redefinitions.

The (0,2) mirror map can then be summarized as follows
\begin{align}
\label{eq:02mirrmap}
\Delta &\leftrightarrow\Delta^\circ~,	&\kappa_a&\leftrightarrow\kappah_{\ah}~,		&b_{m\rho} &\leftrightarrow \bh_{\rho m}~,
\end{align}
where $\bh_{\rho m}=(b_{m\rho})^\top$ denotes the transpose of the matrix $b_{m\rho}$.

It has already been shown that the map \eqref{eq:02mirrmap} passes some significant tests. 
First, the dimensions of the moduli spaces of the theory and its mirror coincide.
Second, the map correctly exchanges the principal components of the singular loci of the A/2- and B/2-twisted theories with those of the mirror. 
While these tests are certainly non-trivial, it is desirable to show that the local observables of the A/2 and B/2 models get exchanged
under the map and more generally to show an equivalence at the level of the correlators. This is what we will present in the rest of the work.

\section{A reflexively plain model}
\label{s:quinticorb}

As recalled above, the mirror map is particularly suggestive for reflexively plain models, as the map exchanges the
entire GLSM moduli spaces. We therefore begin with an example of such a class of models. 

We start with the pair of reflexively plain polytopes
\begin{align}
\Delta&:\begin{pmatrix}
1	&0	&2	&3	&-6\\
0	&1	&4	&3	&-8\\
0	&0	&5	&0	&-5\\
0	&0	&0	&5	&-5
\end{pmatrix}~,
&\Delta^\circ&:\begin{pmatrix}
-1	&-1	&1	&1\\
-1	&-1	&1	&2\\
-1	&-1	&2	&1\\
-1	&4	&-3	&-2\\
4	&-1	&-1	&-2
\end{pmatrix}~,
\end{align}
which, following the discussion in section \ref{s:setup}, completely specify the model.
In particular, we have that
\begin{align}
h^{11}(M)&=1~,			&h^{11}(M^\circ)&=21~,
\end{align}
and we find it more manageable to study the A/2 model for $M$ and the B/2 model for $M^\circ$.
We start with the former.

\subsection{The $M$ model}

It turns out that $\Delta^\circ$ contains no other non-zero lattice point, and the associated fan determines 
the toric variety $V=\P^4/\Z_5$, where the action of the $\Z_5$ quotient on the homogeneous coordinates
of $\P^4$ is given by
\begin{align}
\label{eq:Z5action}
[X_1:X_2:X_3:X_4:X_5]\rightarrow [X_1:\zeta X_2: \zeta^2 X_3:\zeta^3 X_4:\zeta^4 X_5 ]~,
\end{align}
where $\zeta=e^{\frac{2\pi i}5}$. The hypersurface $P$ is presented in terms of a homogeneous polynomial of degree 5
invariant under the above symmetry
\begin{align}
P = \alpha_1 X_1^5+ \cdots +\alpha_5 X_5^5 + \alpha_0 X_1\cdots X_5 + \cdots~.
\end{align}
This has 26 terms, corresponding to the lattice points of $\Delta$.
In particular, although the $\Z_5$-action \eqref{eq:Z5action} has 5 fixed points
\begin{align}
&[1:0:0:0:0]~,	&&[0:1:0:0:0]~,	&&[0:0:1:0:0]~,	&&[0:0:0:1:0]~,		&&[0:0:0:0:1]~,
\end{align}
it is not hard to see that a generic hypersurface $P$ will miss these. Thus, \eqref{eq:Z5action} acts freely and 
the corresponding CY hypersurface is smooth. 

The GLSM realizing this geometry is described in terms of the same field content as the usual one-parameter model for the quintic
hypersurface in $\P^4$, 
with $\GU(1)$ gauge charges
\begin{align}
\xymatrix@R=0mm@C=3mm{
X^0	&X^1	&X^2	&X^3	&X^4	&X^5	&\text{F.I.}\\	
-5	&1		&1		&1		&1		&1		&r~,
}
\end{align}
supplemented by the additional quotient $H=\Z_5$ defined by \eqref{eq:Z5action}.
We now turn to the description of the (0,2) deformations of the model. Invariance under \eqref{eq:Z5action}
implies
\begin{align}
\label{eq:quintEdefs}
\cDb \Gamma^0 &=\Sigma \delta X_0~,		&\cDb \Gamma^\rho&= \Sigma e^\rho X_\rho~,
\end{align}
and the $E$-parameters are simply proportional to their (2,2) form, in which case $\delta=-5$ and $e^\rho=1$. 
The $J$-deformations are given by (without considering the $E\cdot J=0$ constraints)
\begin{align}
\label{eq:Jsquintic}
J_\rho&= j_{\rho\rho} X_1^4 + j_{0\rho} \prod_{\rho'\neq\rho}X_{\rho'} +12 \text{ terms}~,
\end{align}
and we count $14\times 5=70$ parameters, which agrees with the number in \cite{Kreuzer:2010ph}.
In \eqref{eq:Jsquintic} we have assumed a specific ordering of the lattice points of $\Delta$,
and we will stick to this choice throughout the rest of this section. Finally, 
according to \eqref{eq:cxpkaehdef}, the invariant K\"ahler coordinate is given by
\begin{align}
\kappa = q \frac{j_{01}\cdots j_{05}}{\alpha_0^5}~.
\end{align}

The phase structure of this model is very simple. At $r>0$ we recover the NLSM on $\cE\rightarrow M$, 
where $M=\{P=0\}$ and $\cE$ is specified by \eqref{eq:bundlecohom}. At $r<0$ instead the field $X_0$ assumes a non-zero
vev and we find a Landau-Ginzburg phase where the orbifold is $\Z_5\times\Z_5$.

Next, we need to determine $\gamma^\rho$, which we recall are specified by the (0,2) supersymmetry constraints 
to span the kernel of the $25\times5$ matrix $b_{m\rho}$, which,
following our discussion above, must have rank $d=4$.
Thus, for the purpose of determining $\gamma^\rho$, we can consider any $5\times5$ minor of $b_{m\rho}$.
We choose this submatrix to be determined by the elements associated to $j_{\rho\rho}$ in \eqref{eq:Jsquintic}, that is
\begin{align}
\label{eq:quintsubmatr}
\begin{pmatrix}
b_{11}	&-1		&-1		&-1		&-1\\
-1		&b_{22}	&-1		&-1		&-1\\
-1		&-1		&b_{33}	&-1		&-1\\
-1		&-1		&-1		&b_{44}	&-1\\
-1		&-1		&-1		&-1		&b_{55}
\end{pmatrix}~.
\end{align}
This matrix has rank at most 4 when the relation 
\begin{align}
\label{eq:quinticb5cond}
b_{55}+1&=\frac{(b_{44}+1)(b_{33}+1)(b_{22}+1)(b_{11}+1)}{b_{11}b_{22}b_{33}b_{44}-b_{11}b_{22}-b_{11}b_{33}-b_{11}b_{44}-b_{22}b_{33}-b_{22}b_{44}-b_{33}b_{44}-2b_{11}-2b_{22}-2b_{33}-2b_{44}-3}
\end{align}
holds. 
The (0,2) supersymmetry constraints then read
\begin{align}
\label{eq:gambconstrs}
b_{11}\gamma^1 - \gamma^2-\gamma^3-\gamma^4-\gamma^5=0~,\nonumber\\
-\gamma^1 +b_{22} \gamma^2-\gamma^3-\gamma^4-\gamma^5=0~,\nonumber\\
-\gamma^1 - \gamma^2+b_{33}\gamma^3-\gamma^4-\gamma^5=0~,\nonumber\\
-\gamma^1 - \gamma^2-\gamma^3+b_{44}\gamma^4-\gamma^5=0~,\nonumber\\
-\gamma^1 - \gamma^2-\gamma^3-\gamma^4+b_{55}\gamma^5=0~.
\end{align}
We can interpret four of these equations as determining $\gamma^\rho$ up to a rescaling, which 
can then be uniquely specified by the remaining $\GL(1,\C)$ field redefinition. 
However, it turns out that the solution will look prettier if we allow for some redundancy. 
In fact, a solution to \eqref{eq:gambconstrs} is given by
\begin{align}
\label{eq:quinticgammas}
\gamma^1&= \frac{b_{44}+1}{b_{11}+1}\gamma^4~,
&\gamma^2 &= \frac{b_{44}+1}{b_{22}+1}\gamma^4~, 
&\gamma^3&= \frac{b_{44}+1}{b_{33}+1}\gamma^4~, 
&\gamma_5&=\frac{b_{44}+1}{b_{55}+1}\gamma^4~,
\end{align}
and we can make use of the $\GL(1,\C)$ rescaling to set $\gamma^4=1$. 
As advertised, these parameters are not independent, but are related by \eqref{eq:quinticb5cond}.
This is just a overly simple manifestation of the fact, mentioned above, that the mirror map assumes a more natural
form in an overdetermined parameter space. For all practical purposes, we can carry on as if $b_{11},\dots,b_{55}$ were independent parameters
and impose \eqref{eq:quinticb5cond} only after actual calculations are performed.
Finally, $\delta$ assumes the form 
\begin{align}
\label{eq:deltaquint}
\delta =-1-b_{44}~,
\end{align}
where we used \eqref{eq:quinticb5cond}.

\subsubsection{A/2 correlators}

In order to solve the A/2-twisted $M$ model we employ the strategy developed in \cite{Morrison:1994fr} for (2,2) theories and later
extended to a class of (0,2) linear models in \cite{McOrist:2008ji}. 
Briefly, the idea is that correlators of the A/2-twisted $M$ model can be extracted 
from the correlators of the A/2-twisted $V$ model by applying the (0,2) version of the quantum restriction formula.
These latter correlators, in turn, are much easier to compute as one can rely on the power
of toric geometry techniques.
Thus, we begin by solving the A/2-twisted $V$ model.

The A/2-twisted $V$ model is completely determined by the chirality conditions \eqref{eq:quintEdefs}, 
which we express in matrix form as
\begin{align}
\cDb \Gamma^\rho &= \Mt(\Sigma)^{\rho\rho'} X_{\rho'}~,		&\Mt(\Sigma)&=\diag(e^1\Sigma,e^2\Sigma, e^3\Sigma,e^4\Sigma,e^5\Sigma)~,
\end{align}
where $\rho,\rho'=1,\dots,5$. The V model can be easily solved with Coulomb branch techniques \cite{McOrist:2007kp}. The effective superpotential for $\sigma$
-- the lowest component in the multiplet $\Sigma$ -- on the Coulomb branch is given by
\begin{align}
\Jt=\log\left[ q^{-1} \det \Mt \right] = \log\left[ q^{-1} \left(\prod_{\rho=1}^5 e^\rho\right) \sigma^5 \right]~.
\end{align} 
The Coulomb branch vacua are solutions to $\Jt=0$, leading to the quantum cohomology relations
\begin{align}
\label{eq:02quintqcr}
\left(\prod_{\rho=1}^5 \gamma^\rho \right)\sigma^5 = \kappa~,
\end{align}
which we expressed in terms of the invariant coordinates.
Thus, the generic $V$ model correlator is given by
\begin{align}
\la \sigma^{4+5k}\ra 
&=\left( \prod_{\rho=1}^5 e^\rho \right)^{-1}\left( \prod_{\rho=1}^5 \gamma^\rho \right)^{-k}\kappa^k~.
\end{align}
It is natural to normalize the correlators such that
\begin{align}
\la \sigma^{4+5k}\ra &=\frac15 \left({ \kappa \over \prod_{\rho=1}^5 \gamma^\rho} \right)^k~,
\end{align}
where the extra factor of $5^{-1}$ follows from the quotient by $H=\Z_5$ \cite{Morrison:1994fr}.
Next, we turn to the A/2-twisted $M$ model, and we make use of the
quantum restriction formula, that for this example reads
\begin{align}
\llangle \sigma^k \rrangle =\la \sigma^k \frac{-\delta\sigma}{1-\delta^5\sigma^5}\ra~.
\end{align}
In particular, it follows from \eqref{eq:02quintqcr} that
\begin{align}
\delta^5\sigma^5&=\frac{\delta^5 }{\prod_{\rho=1}^5 \gamma^\rho}\kappa~,
\end{align}
and therefore
\begin{align}
\llangle \sigma^3 \rrangle =\frac15 {-\delta \over 1- \frac{\delta^5 }{\prod_{\rho=1}^5 \gamma^\rho}\kappa}~.
\end{align}
This formula passes two important checks. First, it reduces to the known expression on the (2,2) locus.
Second, the correlators are singular where the denominator vanishes, and this reproduces
the formula for the principal component of the discriminant locus of the A/2-twisted theory \cite{Melnikov:2010sa}.

Finally, we substitute the expressions \eqref{eq:quinticgammas} and \eqref{eq:deltaquint} for $\gamma^\rho$ and $\delta$
into our correlators and we obtain
\begin{align}
\llangle \sigma^3 \rrangle
&=\frac15\frac{b_{44}+1}{1+(b_{11}+1)(b_{22}+1)(b_{33}+1)(b_{44}+1)(b_{55}+1)\kappa}~.
\end{align}
The natural observables of the A/2 model are represented by 
\begin{align}
H_0&=\delta\sigma~,		&H_\rho &= \gamma^\rho \sigma~,\qquad \rho=1,\dots,5~.
\end{align}
It is now simple to compute the full list of correlators
\begin{align}
\label{eq:A2quintcorrs}
\llangle H_{i} H_{j} H_{k} \rrangle &=\frac{\beta_{44}^4}{\beta_{ii}\beta_{jj}\beta_{kk}}\frac{5^{-1}}{1+\kappa \prod_\rho \beta_{\rho\rho}}~,\nonumber\\
\llangle H_{i} H_{j} H_0 \rrangle &=-\frac{\beta_{44}^4}{\beta_{ii}\beta_{jj}}\frac{5^{-1}}{1+\kappa\prod_\rho \beta_{\rho\rho}}~,\nonumber\\
\llangle H_{i}  H_0^2 \rrangle &=\frac{\beta_{44}^4}{\beta_{ii}}\frac{5^{-1}}{1+\kappa \prod_\rho \beta_{\rho\rho}}~,\nonumber\\
\llangle H_0^3 \rrangle &=-\frac{5^{-1}\beta_{44}^4}{1+\kappa \prod_\rho \beta_{\rho\rho}}~,
\end{align}
where $i,j,k=1,\dots,5$ and we introduced $\beta_{m\rho}\equiv b_{m\rho}+1$.

\subsection{The $M^\circ$ model}

We now tackle the B/2-twisted theory for the mirror model,
the corresponding GLSM being a fairly untreatable 21-parameter model. 
As sketched above, we can still solve the model using the following two facts. 
First, it is known that the B/2-twisted model does not receive 
worldsheet instanton corrections \cite{McOrist:2008ji}. 
This implies that we can perform the computation at any point in the K\"ahler moduli space. 
Second, one of the phases of the linear model admits a Landau-Ginzburg orbifold description. Thus,
we can make use of the explicit LGO theory to completely solve the full B/2 model.

The $M^\circ$ LG model is described in terms of the chiral superfields $Y_1,\dots,Y_5$, 
to which we assign R-charge $\frac15$, and we take the orbifold by $\Z_5^3$, generated by
\begin{align}
\label{eq:Z52actions}
\GU(1)_0&: [Y_1:Y_2:Y_3:Y_4:Y_5]\rightarrow [\zeta Y_1:\zeta Y_2: \zeta Y_3:\zeta Y_4: \zeta Y_5 ]~,\nonumber\\
\GU(1)_1&: [Y_1:Y_2:Y_3:Y_4:Y_5]\rightarrow [Y_1:\zeta Y_2: \zeta^3 Y_3:\zeta Y_4: Y_5 ]~,\nonumber\\
\GU(1)_2&: [Y_1:Y_2:Y_3:Y_4:Y_5]\rightarrow [Y_1:\zeta Y_2: \zeta Y_3: Y_4: \zeta^3Y_5 ]~,
\end{align}
where $\zeta=e^{\frac{2\pi i}5}$. 
The most general polynomial of degree 5 invariant under \eqref{eq:Z52actions} is given by
\begin{align}
P^\circ&=\alphah_1 Y_1^5 +\alphah_2 Y_2^5+\alphah_3 Y_3^5+\alphah_4 Y_4^5  + \alphah_5 Y_5^5 + \alphah_0 Y_1Y_2Y_3Y_4Y_5~,
\end{align}
and the unique invariant complex structure coordinate reads
\begin{align}
\kappah = \frac{\alphah_1 \cdots \alphah_5}{\alphah_0^5}~.	
\end{align}
Lastly, the (0,2) superpotential is determined by
\begin{align}
J^\circ_m &=\frac{\jh_{0m}\alphah_m}{\alphah_0}(\bh_{mm}+1) Y_m^4 +\jh_{0m} \prod_{m'\neq m} Y_{m'}~,		&m,m'&=1,\dots,5~.
\end{align}
A few comments are in order here. First, our notation is such that
$\bh_{mm}$ is to be understood
as $\bh_{m,\rho=m}=(b_{\rho,m=\rho})^\top$, $\rho=1,\dots,5$.
Second, although in the full GLSM the index $m$ assumes the values 
$m=1,\dots,26$, only a subset of the corresponding variables $Y_m$ actively plays a role in the LG phase, 
the other coordinates being massive in the limit and we can integrate them out.
In particular, our choice of basis for the matrix $b$ in the original model 
is such that these correspond to 
$m=1,\dots,5$.

\subsubsection{B/2 correlators}

The natural observables in the B/2-twisted LGO theory are
\begin{align}	
\muh_0&=\alphah_0 Y_1 Y_2Y_3Y_4Y_5~,		&\muh_m&=\alphah_m Y_m^5~,	\qquad\quad m=1,\dots,5~.
\end{align}
The formula for B/2-twisted LG correlators \cite{Vafa:1990mu,Melnikov:2007xi,Melnikov:2009nh} is fairly straightforward to implement
and is expressible in terms of a local Grothendieck residue
\begin{align}
\label{eq:02LGOcorrs}
\la \muh_{m_1}\muh_{m_2}\muh_{m_2} \ra_{\text{B/2}}= \frac1{(2\pi i)^5}\int_\Gamma dY_1\wedge\dots\wedge dY_5 \frac{\muh_{m_1}\muh_{m_2}\muh_{m_3}}{J^\circ_1\cdots J^\circ_5}~,
\end{align}
where $m_{1,2,3}=0,\dots,5$, and the integral is determined in terms of a cycle $\Gamma=\{Y|\ |J^\circ_m|^2 > \ep_m  \}$, $\ep_{m}>0$. 
Explicitly, such integrals can be evaluated by taking advantage of the transformation law for local residues \cite{Griffiths:1978pa}. 
As these techniques are quite standard,\footnote{For a review, the reader can refer to \cite{Bertolini:2018now}.} 
we simply present the full list of correlators
\begin{align}
\label{eq:B2quintcorrs}
\la \muh_{i}\muh_{j}\muh_{k} \ra_{\text{B/2}}&=-\frac{\alphah_0^3}{\prod_m \jh_{0m}}\frac1{\betah_{ii}\betah_{jj}\betah_{kk}} \frac1{1+\kappah\prod_{m}\betah_{mm}}~,\nonumber\\
\la \muh_{i}\muh_{j}\muh_0 \ra_{\text{B/2}}&=\frac{\alphah_0^3}{\prod_m \jh_{0m}}\frac1{\betah_{ii}\betah_{jj}}\frac1{1+\kappah\prod_{m}\betah_{mm}}~,\nonumber\\
\la \muh_{i}\muh_0^2 \ra_{\text{B/2}}&=-\frac{\alphah_0^3}{\prod_m \jh_{0m}}\frac1{\bh_{ii}} \frac1{1+\kappah\prod_{m}\betah_{mm}}~,\nonumber\\
\la \muh_0^3 \ra_{\text{B/2}}&= \frac{\alphah_0^3}{\prod_m \jh_{0m}}\frac1{1+\kappah\prod_{m}\betah_{mm}}~,
\end{align}
where $i,j,k=1,\dots,5$ and we defined $\betah_{\rho m}\equiv \bh_{\rho m}+1$. 
The formula \eqref{eq:02LGOcorrs} provides unrenormalized correlators, and we do not have an independent way
to determine the normalization. 
Nonetheless, it is natural to normalize the correlators as follows
\begin{align}
\label{eq:B2natnorm}
\la \muh_1\muh_2\muh_3 \ra_{\text{B/2,N}}&=\frac{\prod_{\rhoh}\jh_{0,\rhoh}}{\alpha_0^3} \la \muh_1\muh_2\muh_3 \ra_{\text{B/2}}~,
\end{align}
so that the result depends explicitly only on invariant quantities.
By comparing these with the A/2 correlators in \eqref{eq:A2quintcorrs} we find 
the relation
\begin{align}
\llangle H_{m_1}H_{m_2} H_{m_3} \rrangle \quad \leftrightarrow \quad -\frac15(\bh_{44}+1)^4 \la \muh_{m_1}\muh_{m_2}\muh_{m_3} \ra_{\text{B/2,N}}~.
\end{align}
Thus, we find complete agreement, up to the relative normalization, between the two sets of correlators 
under the mirror map 
if the observables are exchanged according to $\muh_{m}\leftrightarrow H_{m}$. 

Notice that $b_{44}=-1$ corresponds to a singularity of the theory. For example, in the LGO description, 
at this locus the ideal $\la J^\circ\ra$ fails to be zero-dimensional.
Thus, the relative normalization between the correlators 
is not allowed to vanish and can be absorbed, for instance, through a non-singular redefinition of the fields. 

\section{Mirror subfamilies in a non-reflexively plain model}
\label{s:octic}

In this section we study the mirror map in a non-reflexively plain model, namely the two-parameter model describing
in its large radius phase an octic hypersurface in the toric resolution of the weighted projective space $\P^4_{11222}$ \cite{Morrison:1994fr}.

It has been shown that the full linear model for this example is not mirror symmetric \cite{Kreuzer:2010ph}, but there
exist subfamilies of the model and its mirror, identified by the restriction to the ``diagonal" form of the $E$-parameters \eqref{eq:Esdiagform},
which are conjectured to be exchanged by the mirror map \cite{Melnikov:2010sa}. In this section we will show that this is
the case at the level of the correlators.

\subsection{The $M$ model}

The model is specified by seven chiral multiplets $X_0, X_\rho$, $\rho=1,\dots,6$, coupled to the $\GU(1)^{\oplus2}$ gauge group via
\begin{align}
\label{eq:octicgaugechs}
\xymatrix@R=0mm@C=3mm{
			&X_0		&X_1		&X_2	&X_3	&X_4	&X_5	&X_6		&\text{F.I.}\\
\GU(1)_1		&-4			&1			&1		&1		&0		&0		&1			&r_1\\		
\GU(1)_2		&0			&0			&0		&0		&1		&1		&-2			&r_2
}
\end{align}
The polynomial defining the octic hypersurface assumes the form
\begin{align}
P&=\alpha_1 X_1^4+\alpha_2X_2^4+\alpha_3X_3^4+\alpha_4 X_4^8X_6^4+\alpha_5 X_5^8X_6^4+\alpha_6X_4^4X_5^4X_6^4+\alpha_0 X_1X_2X_3X_4X_5X_6+\cdots~.
\end{align}

The left-moving fermions $\Gamma^0,\Gamma^\rho$ couple to the gauge group with the same charges as in \eqref{eq:octicgaugechs} and have chirality determined by 
\begin{align}
\cDb \Gamma^0 &= X_0 \bSigma\cdot\bdelta~,				&\cDb \Gamma^\rho = X_\rho \bSigma\cdot \be^\rho~,
\end{align}
where we have introduced the vectors
\begin{align}
\bdelta&=\begin{pmatrix} \delta_1 \\ \delta_2 \end{pmatrix}~,	&\be^\rho&=\begin{pmatrix} e^\rho_1 \\ e^\rho_2 \end{pmatrix}~,		
&\bSigma&=\begin{pmatrix} \Sigma_1 & \Sigma_2 \end{pmatrix}~.
\end{align}
The $J$-deformations take the following form
\begin{align}
\label{eq:octicJdefs}
J_\rho&=j_{\rho\rho}X_\rho^3+j_{0\rho} \prod_{\rho'\neq\rho} X_{\rho'}+\cdots~,		&\rho&=1,2,3~,\nonumber\\
J_\rho&=j_{\rho\rho}X_\rho^7X_6^4+j_{6\rho}X_\rho^3X_6^4\prod_{\substack{\rho'=4,5\\\rho'\neq\rho}}X_{\rho'}^4+j_{0\rho} \prod_{\rho'\neq\rho} X_{\rho'}+\cdots~,		&\rho&=4,5~,\nonumber\\
J_6&=j_{66}X_4^4X_5^4X_6^3+j_{46}X_4^8X_6^3+j_{56}X_5^8X_6^3+j_{06} \prod_{\rho'\neq6} X_{\rho'}+\cdots~,		
\end{align}
and the invariant K\"ahler coordinates are
\begin{align}
\kappa_1&=q_1\frac{j_{01}j_{02}j_{03}j_{06}}{\alpha_0^4}~,		&\kappa_2&=q_2\frac{j_{04}j_{05}}{j_{06}^2}~.
\end{align}
The matrix $b_{m\rho}$ for this example has dimension $104\times6$, and it would be pointless to write it down in its entirety. 
We instead assume an ordering of the lattice points in $\Delta$ (or equivalently, an ordering of the rows of $b_{m\rho}$) 
corresponding to the choice \eqref{eq:octicJdefs}, and consider the $6\times6$ minor
\begin{align}
\label{eq:bmatrixoctic}
\begin{pmatrix}
b_{11} 	&-1		&-1		&-1		&-1		&-1\\
-1		&b_{22} 	&-1		&-1		&-1		&-1\\
-1		&-1		&b_{33} 	&-1		&-1		&-1\\
-1		&-1		&-1 		&b_{44}	&-1		&b_{46}\\
-1		&-1		&-1 		&-1		&b_{55}	&b_{56}\\
-1		&-1		&-1 		&b_{64}	&b_{65}	&b_{66}
\end{pmatrix}~.
\end{align}
For generic values of the parameters this matrix has full rank. However, when the parameters satisfy the relations
\begin{align}
\label{eq:octbconstr}
b_{55}+1&=\frac{(b_{44}+1)(b_{33}+1)(b_{22}+1)(b_{11}+1)}{b_{11}b_{22}b_{33}b_{44}-b_{11}b_{22}-b_{11}b_{33}-b_{11}b_{44}-b_{22}b_{33}-b_{22}b_{44}-b_{33}b_{44}-2b_{11}-2b_{22}-2b_{33}-2b_{44}-3}~,\nonumber\\
b_{66}&=\frac{b_{44}b_{56}+b_{46}b_{64}-b_{56}b_{64}+b_{46}}{b_{44}+1}~,\nonumber\\
b_{65}&=\frac{b_{44}b_{55}-b_{55}b_{64}-b_{64}-1}{b_{44}+1}~,\nonumber\\
b_{46}&=-\frac1{(b_{33}+1)(b_{22}+1)(b_{11}+1)}\Big[ b_{11}b_{22}b_{33}b_{44}b_{56}-b_{11}b_{22}b_{44}-b_{11}b_{22}b_{56}-b_{11}b_{33}b_{44} \nonumber\\
&\qquad\qquad -b_{11}b_{33}b_{56}-b_{11}b_{44}b_{56}-b_{22}b_{33}b_{44}-b_{22}b_{33}b_{56}-b_{22}b_{44}b_{56}-b_{33}b_{44}b_{56}-b_{11}b_{22}\nonumber\\
&\qquad\qquad-b_{11}b_{33}-2b_{11}b_{44}-2b_{11}b_{56}-b_{22}b_{33}-2b_{22}b_{44}-2b_{22}b_{56}-2b_{33}b_{44}-2b_{33}b_{56}\nonumber\\
&\qquad\qquad-2b_{44}b_{56}-2b_{11}-2b_{22}-2b_{33}-3b_{44}-3b_{56}-3\Big]~,
\end{align}
the rank of \eqref{eq:bmatrixoctic} reduces to at most 4. We will however assume that the remaining parameters are generic enough such that
the rank of \eqref{eq:bmatrixoctic} will be exactly 4. Thus, of the 10 parameters in \eqref{eq:bmatrixoctic} only 6 are independent.\footnote{This is consistent with the
fact that the subvariety of $p\times q$ matrices of rank at least $r$ in the variety of $p\times q$ matrices has codimension $(p-r)(q-r)$.} 
As in the previous example, the solution
takes a much prettier form if we again allow for some redundancy. In particular, it is convenient to 
eliminate the explicit dependence of $b_{65}$ and $b_{66}$ through the constraints \eqref{eq:octbconstr}, and regarding all the other quantities as ``independent" parameters. 
With this strategy, the kernel of \eqref{eq:bmatrixoctic} assumes a simple form
\begin{align}
\label{eq:02octicgammas}
\bgamma&=\begin{pmatrix}
\frac{b_{56}+1}{b_{11}+1}		&\frac{b_{56}+1}{b_{22}+1}	&\frac{b_{56}+1}{b_{33}+1}		&\frac{b_{56}-b_{46}}{b_{44}+1}			&0							&1\\
0						&0						&0							&2\frac{b_{46}+1}{b_{44}+1}				&2\frac{b_{56}+1}{b_{55}+1}		&-2
\end{pmatrix}~,		
&\bdelta&=\begin{pmatrix}
-(1+b_{56})\\0
\end{pmatrix}~,
\end{align}
where we used the $\GL(2,\C)$ remaining field redefinitions to set, in particular, $\gamma^1_2=0$ and $\bgamma^6$ to its (2,2) value. Notice that $b_{64}$ does not enter our solution,
neither explicitly nor implicitly through \eqref{eq:octbconstr}. Therefore the correlators will depend on 5 parameters, 
instead of the naive 6. We will see when we study the mirror
that this is the correct behavior.

\subsubsection{A/2 correlators}

At this point everything is in place, and we can proceed to solve the model. 
As in the previous example, we first tackle the A/2-twisted $V$ model and then derive the $M$ model
correlators by taking advantage of the quantum restriction formula. 

Let us recast the chirality conditions for the left-moving Fermi fields in matrix notation
\begin{align}
\cDb \bGamma^{(1)} &= \bX_{(1)} \cdot \Mt_1~, 	&\cDb \bGamma^{(2)} &= \bX_{(2)} \cdot \Mt_2~,		&\cDb \bGamma^{(3)} &= \bX_{(3)} \cdot \Mt_3~,  
\end{align}
where we organized the fields according to their gauge charges
\begin{align}
\bX_{(1)} &= \begin{pmatrix} X_1 &X_2&X_3\end{pmatrix}~,	&\bX_{(2)} &= \begin{pmatrix} X_4 &X_5\end{pmatrix}~, &\bX_{(3)} &= X_6 ~,\nonumber\\
\bGamma^{(1)} &= \begin{pmatrix} \Gamma^1 \\\Gamma^2\\\Gamma^3\end{pmatrix}~,	
&\bGamma^{(2)} &= \begin{pmatrix} \Gamma^4 \\\Gamma^5 \end{pmatrix}~, &\bGamma^{(3)} &=  \Gamma^6 ~.
\end{align}
Given the form \eqref{eq:02octicgammas} of $\bgamma^\rho$, the corresponding $E$-parameters are 
described by the matrices
\begin{align}
\Mt_1&=\begin{pmatrix}
e^1\Sigma_1	&0	&0\\
0			&e^2\Sigma_1	&0\\
0			&			&e^3 \Sigma_1
\end{pmatrix}~,
&\Mt_2&=\begin{pmatrix}
e^4_1\Sigma_1+e^4_2\Sigma_2	&0\\
0			&e^5\Sigma_2
\end{pmatrix}~,
&\Mt_3&=\begin{pmatrix}
e^6_1\Sigma_1	+e^6_2\Sigma_2
\end{pmatrix}~,
\end{align}
and correspondingly we define the vectors
\begin{align}
\be^{1,2,3}&=\begin{pmatrix} e^{1,2,3}\\0\end{pmatrix}~,		&\be^4&=\begin{pmatrix} e^4_1\\e^4_2\end{pmatrix}~,
&\be^5&=\begin{pmatrix} 0\\e^5\end{pmatrix}~,				&\be^6&=\begin{pmatrix} e^6_1\\e^6_2\end{pmatrix}~.
\end{align}
Next, we construct the effective potential for the $\sigma$ fields in the Coulomb branch
\begin{align}
\label{eq:octeffpot}
\Jt_1&=\log\left[ q_1^{-1} \det \Mt_1 \det \Mt_3 \right]~,		&\Jt_2&=\log\left[ q_2^{-1} \det \Mt_2 (\det \Mt_3)^{-2} \right]~,
\end{align}
and setting $\Jt_1=\Jt_2=0$ we obtain the quantum cohomology relations
\begin{align}
\label{eq:octicqcr}
e^1e^2e^3\sigma_1^3(e^6_1\sigma_1+e^6_2\sigma_2)&=q_1~,	&(e^4_1\sigma_1+e^4_2\sigma_2)e^5\sigma_2&=(e^6_1\sigma_1+e^6_2\sigma_2)^2q_2~.
\end{align}
The formula that yields the correlators reads
\begin{align}
\la \sigma_1^a \sigma_2^b \ra = \sum_{\sigma|\Jt=0}\sigma_1^a \sigma_2^b\left[ \det\Jt_{\mu,\nu}\det \Mt_1 \det \Mt_2 \det \Mt_3 \right]^{-1}~,
\end{align}
where $a+b=4t$. Here $\mu,\nu=1,2$ and $\Jt_{\mu,\nu}$ indicates the matrix of derivatives of \eqref{eq:octeffpot},
and the full measure reads
\begin{align}
 \det\Jt_{\mu,\nu}\det \Mt_1 \det \Mt_2 \det \Mt_3=4(2e^6_1e^4_2\sigma_2+e^6_1e^4_1\sigma_1-e^6_2e^4_1\sigma_2)e^1e^2e^3e^5 \sigma_1^3~.
\end{align}
It turns out to be more convenient to solve the correlators in terms of $\sigma_1$ and of the ratio $z=\sigma_2/\sigma_1$,
in terms of which the relations \eqref{eq:octicqcr} assume the form
\begin{align}
\sigma_1^4 &= \frac{\kappa_1}{\gamma^1\gamma^2\gamma^3(\gamma^6_1+\gamma^6_2 z)}~,		
&(\gamma^4_1+\gamma^4_2z)\gamma^5z&=(\gamma^6_1+\gamma^6_2z)^2\kappa_2~.
\end{align}
Let us define 
\begin{align}
T(z)=(\gamma^4_1+\gamma^4_2z)\gamma^5z-(\gamma^6_1+\gamma^6_2z)^2\kappa_2~,
\end{align}
then we have
\begin{align}
\la \sigma_1^a \sigma_2^{4t-a} \ra 
&= \frac{\kappa_1^{t-1}}{4e^1e^2e^3e^5}\sum_{z|T(z)=0}\frac{z^{4t-a}}{\left(\gamma^1\gamma^2\gamma^3(\gamma^6_1+\gamma^6_2 z)\right)^{t-1}(2e^6_1e^4_2z+e^6_1e^4_1-e^6_2e^4_1z)}~.
\end{align}
We choose to normalize the correlators as follows 
\begin{align}
\label{eq:octcorrnorm}
\la \sigma_1^a \sigma_2^{4t-a} \ra &= \kappa_1^{t-1}\sum_{z|T(z)=0}\frac{z^{4t-a}(e^6_1e^4_2-e^6_2e^4_1)}{\left(\gamma^1\gamma^2\gamma^3(\gamma^6_1+\gamma^6_2 z)\right)^{t-1}(2e^6_1e^4_2z+e^6_1e^4_1-e^6_2e^4_1z)}~.
\end{align}
It is not hard to evaluate this expression for any $a,t$. As an example we present the correlators for $t=1$, which read
\begin{align}
\la \sigma_1^4\ra &=-\frac{\gamma^6_2}{\gamma^6_1}~,\nonumber\\
\la \sigma_1^3\sigma_2\ra &=1~,\nonumber\\
\la \sigma_1^2\sigma_2^2\ra &=\frac{\gamma^6_1\gamma^6_2\kappa_2-\gamma^4_1\gamma^5}{-(\gamma^6_2)^2\kappa_2+\gamma^4_2\gamma^5}~,\nonumber\\
\la \sigma_1\sigma_2^3\ra &=\frac{(\gamma^6_1\gamma^6_2)^2\kappa_2^2+\left((\gamma^6_1)^2\gamma^4\gamma^5-3\gamma^4_1\gamma^5\gamma^6_1\gamma^6_2\right)\kappa_2
+(\gamma^4_1\gamma^5)^2}{\left(-(\gamma^6_2)^2\kappa_2+\gamma^4_2\gamma^5\right)^2}~.
\end{align}
In particular, our choice of normalization is such that the correlator $\la \sigma_1^3\sigma_2\ra =1$ assumes 
throughout the whole parameter space the same value as on the (2,2) locus.

In order to solve the A/2-twisted M model,
we implement the quantum restriction formula, which for this example reads
\begin{align}
\llangle \sigma_1^a\sigma_2^{3-a} \rrangle &= \la \sigma_1^a\sigma_2^{3-a} \frac{-\bdelta\cdot \bsigma}{1-(\bdelta\cdot \bsigma)^4}\ra~.
\end{align}
Plugging in the solution \eqref{eq:octcorrnorm} and substituting for $\sigma_1^4$, we obtain
\begin{align}
\llangle \sigma_1^a\sigma_2^{3-a} \rrangle 
&=(e^6_1e^4_2-e^6_2e^4_1)  \sum_{z|T(z)=0}\frac{-\delta\gamma^1\gamma^2\gamma^3(\gamma^6_1+\gamma^6_2 z) z^{3-a}}{(\gamma^1\gamma^2\gamma^3(\gamma^6_1+\gamma^6_2 z)-\delta^4\kappa_1)(2e^6_1e^4_2z+e^6_1e^4_1-e^6_2e^4_1z)}~.
\end{align}
For example, we have
\begin{align}
\llangle \sigma_1^3 \rrangle =\frac{\delta (\gamma^1\gamma^2\gamma^3)^2\gamma^5\gamma^6_2(\gamma^4_2\gamma^6_1-\gamma^4_1\gamma^6_2)}{D}~,
\end{align}
where
\begin{align}
D&=-\delta^8(\gamma^6_2)^2\kappa_1^2\kappa_2+\delta^8\gamma^4_2\gamma^5\kappa_1^2-2\delta^4\gamma^1\gamma^2\gamma^3\gamma^4_2\gamma^5\gamma^6_1\kappa_1
+\delta^4\gamma^1\gamma^2\gamma^3\gamma^4_1\gamma^5\gamma^6_2\kappa_1\nonumber\\
&\quad+(\gamma^1\gamma^2\gamma^3)^2\gamma^4_2\gamma^5(\gamma^6_1)^2
-(\gamma^1\gamma^2\gamma^3)^2\gamma^4_1\gamma^5\gamma^6_1\gamma^6_2
\end{align}
is the principal component of the discriminant locus.
Now, plugging in the expressions for the $\bgamma^\rho$ and $\bdelta$ we have
\begin{align}
\label{eq:octiccorrsV1}
\llangle \sigma_1^3 \rrangle &= \frac{2(b_{56}+1)}{D}~,
\end{align}
and
\begin{align}
D&=-\left(\prod_{\rho=1}^5 (b_{\rho\rho}+1)\right)\left(\prod_{\rho'=1}^3 (b_{\rho'\rho'}+1)\right)\kappa_1^2\kappa_2 \nonumber\\
&\quad+\left(\left(\prod_{\rho'=1}^3 (b_{\rho'\rho'}+1)\right)(b_{46}+1)\kappa_1-1\right)\left(\left(\prod_{\rho'=1}^3 (b_{\rho'\rho'}+1)\right)(b_{56}+1)\kappa_1-1\right)~.
\end{align}
The remaining correlators are given by 
\begin{align}
\label{eq:octiccorrsV2}
\llangle \sigma_1^2 \sigma_2 \rrangle &= \frac{(b_{56}+1)(1-\left(\prod_{\rho'=1}^3 (b_{\rho'\rho'}+1)\right)(b_{56}+1)\kappa_1)}{D}~,\nonumber\\
\llangle \sigma_1\sigma_2^2 \rrangle &=\frac{b_{56}+1}{2DD_1} \left[ (b_{56}+1)(b_{46}-b_{56})+
2\left(\prod_{\rho=1}^5 (b_{\rho\rho}+1)\right)(b_{56}+1)\kappa_1\kappa_2\right. \nonumber\\
&\qquad\qquad\qquad \left.-\left(\prod_{\rho'=1}^3 (b_{\rho'\rho'}+1)\right)(b_{46}-b_{56})(b_{56}+1)^2
\kappa_1-(b_{44}+1)(b_{55}+1)\kappa_2\right]~,\nonumber\\
\llangle \sigma_2^3 \rrangle&=\frac{b_{56}+1}{4DD_1^2}\left[ (b_{44}+1)^2(b_{55}+1)^2 \kappa_2^2 
 -3 \left(\prod_{\rho=1}^5 (b_{\rho\rho}+1)\right)(b_{44}+1)(b_{55}+1)(b_{56}+1)\kappa_1\kappa_2^2 \right.\nonumber\\
&\qquad\qquad\qquad +(b_{44}+1)(b_{55}+1)(b_{56}+1)(3b_{56}-2b_{46}+1)\kappa_2 \nonumber\\
&\qquad\qquad\qquad -\left(\prod_{\rho=1}^5 (b_{\rho\rho}+1)\right)(b_{56}+1)^2(4b_{56}-3b_{46}+1)\kappa_1\kappa_2\nonumber\\
&\qquad\qquad\qquad \left. -(b_{56}+1)^2(b_{56}-b_{46})^2\left(\left(\prod_{\rho'=1}^3 (b_{\rho'\rho'}+1)\right)(b_{56}+1)\kappa_1-1\right) \right]~,
\end{align}
where 
\begin{align}
D_1&=(b_{46}+1)(b_{56}+1)-(b_{44}+1)(b_{55}+1)\kappa_2
\end{align}
is an additional component of the discriminant locus.

Now, the observables of the A/2 model are
\begin{align}
H_\rho&=\bgamma^\rho \cdot \bsigma~,		&H_0&=\bdelta \cdot \bsigma~.
\end{align}
Explicitly, by implementing our solution \eqref{eq:02octicgammas}, these assume the expressions
\begin{align}
H_1&=\frac{b_{56}+1}{b_{11}+1} \sigma_1~,		&H_2&=\frac{b_{56}+1}{b_{22}+1} \sigma_1~,		&H_3&=\frac{b_{56}+1}{b_{33}+1} \sigma_1~,\nonumber\\
H_4&=\frac{b_{56}-b_{46}}{b_{44}+1} \sigma_1+2\frac{b_{46}+1}{b_{44}+1}\sigma_2~,	
&H_5&=2\frac{b_{56}+1}{b_{55}+1}\sigma_2~,		&H_6&=\sigma_1-2\sigma_2~,
\end{align}
and 
\begin{align}
H_0 = -(b_{56}+1) \sigma_1~.
\end{align}
With a bit of algebra we can produce all the correlators, as these are expressed in terms of \eqref{eq:octiccorrsV1} and \eqref{eq:octiccorrsV2}.
As an example
\begin{align}
\label{eq:H63corr}
\llangle H_6^3 \rrangle &= \llangle \sigma_1^3 \rrangle -6\llangle \sigma_1^2\sigma_2 \rrangle +12\llangle \sigma_1\sigma_2^2 \rrangle -8\llangle \sigma_2^3\rrangle
\nonumber\\
&=-\frac{2(b_{56}+1)^4}{DD_1^2}\left[ (b_{46}+1)+(b_{56}+1)-\left(\prod_{\rho=1}^5(b_{\rho\rho}+1)\right) \kappa_1\kappa_2 \right. \nonumber\\
&\qquad\qquad\qquad\quad \left.  -\left(\prod_{\rho'=1}^3(b_{\rho'\rho'}+1)\right)\left( (b_{46}+1)^2 +(b_{46}+1)(b_{56}+1) + (b_{56}+1)^2\right)\kappa_1 \right]~.
\end{align}
We present the full list of the A/2 correlators in appendix \ref{app:corrslistsA2}.

\subsection{The $M^\circ$ model}

We now turn to the B/2-twisted mirror model. Again, we make use of two properties of the $M^\circ$ theory, namely that
the model exhibits a LGO phase, and that the B/2-twisted model is not corrected by worldsheet instantons \cite{McOrist:2008ji}. 
Thus, we are able to solve the B/2 model in the LGO phase. 

In this phase the model is described in terms of coordinates $Y_{1,2,3}$ with R-charge $\frac14$, and coordinates $Y_{4,5}$ with R-charge $\frac18$, supplemented by a
$\Z_8\times\Z_4^{\oplus3}$ orbifold with action
\begin{align}
\label{eq:Z43actions}
\GU(1)_0&: [Y_1:Y_2:Y_3:Y_4:Y_5]\rightarrow [\zeta^2 Y_1:\zeta^2 Y_2: \zeta^2 Y_3:\zeta Y_4: \zeta Y_5 ]~,\nonumber\\
\GU(1)_1&: [Y_1:Y_2:Y_3:Y_4:Y_5]\rightarrow [\zeta^2 Y_1: Y_2:  Y_3: Y_4: \zeta^{-4}Y_5 ]~,\nonumber\\
\GU(1)_2&: [Y_1:Y_2:Y_3:Y_4:Y_5]\rightarrow [Y_1:\zeta^2 Y_2:  Y_3: Y_4: \zeta^{-4}Y_5 ]~,\nonumber\\
\GU(1)_3&: [Y_1:Y_2:Y_3:Y_4:Y_5]\rightarrow [Y_1: Y_2:  \zeta^2 Y_3: Y_4: \zeta^{-4}Y_5 ]~,
\end{align}
where $\zeta=e^{\frac{2\pi i}8}$. The most generic polynomial of R-charge 1 and invariant under \eqref{eq:Z43actions} is given by
\begin{align}
P^\circ= \alphah_1 Y_1^4 + \alphah_2 Y_2^4 + \alphah_3 Y^3 +\alphah_4 Y_4^8+\alphah_5 Y^8 + \alphah_0 Y_1Y_2Y_3Y_4Y_5 + \alphah_6 Y_4^4 Y_5^4~.
\end{align}
The invariant complex structure coordinates are according to \eqref{eq:cxpkaehdef}
\begin{align}
\kappah_1 &= \frac{\alphah_1\alphah_2\alphah_3\alphah_6}{\alphah_0^4}~,			&\kappah_2&=\frac{\alphah_4\alphah_5}{\alphah_6^2}~,
\end{align}
while the $J$-deformations take the form
\begin{align}
\label{eq:02supoctic}
J^\circ_m&=\frac{\jh_{0m}\alphah_m}{\alphah_0}(\bh_{mm+1})\left[ Y_m^3+\frac{\alphah_0}{\alphah_m(\bh_{mm+1})} \prod_{\substack{m'=1\\m'\neq m}}^5 Y_{m'}\right]~,	&m&=1,2,3~,\nonumber\\
J^\circ_m&=\frac{\jh_{0m}\alphah_m}{\alphah_0}(\bh_{mm+1}) \left[ Y_m^7+\frac{\alphah_0}{\alphah_m(\bh_{mm}+1)}\prod_{\substack{m'=1\\m'\neq m}}^5 Y_{m'}
+\frac{\alphah_6(\bh_{6m}+1)}{\alphah_m(\bh_{mm}+1)} Y_m^3\prod_{\substack{m'=4\\m'\neq m}}^5 Y_{m'}^4 \right]~,	&m&=4,5~.
\end{align}
Before we proceed with the computation of the correlators, we point out that already at this level we find an important consistency check 
with the original model. In fact, the (0,2) superpotential \eqref{eq:02supoctic} depends explicitly on 7 parameters $\bh_{mm}$, $m=1,\dots,5$,
as well as $\bh_{64}$ and $\bh_{65}$. This matches precisely the dependence we found in the A/2 model. Once we take into 
consideration the (mirror) relations \eqref{eq:octbconstr}, we find that the above parameters satisfy two relations and 
the B/2 model depends on 5 bundle coordinates.

\subsubsection{B/2 correlators}

The natural observables of the B/2-twisted LGO theory are
\begin{align}
\muh_0&= \alphah_0 Y_1Y_2Y_3Y_4Y_5~,		&\muh_m&=\alphah_m Y_m^4~,	\qquad m=1,2,3~,\nonumber\\
\muh_6&= \alphah_6 Y_4^4Y_5^4~,				&\muh_m&=\alphah_m Y_m^8~,	\qquad m=4,5~. 
\end{align}
Again, we can straightforwardly employ the formula \eqref{eq:02LGOcorrs} for evaluating the relevant cubic correlators. 
For example, we present the correlator 
\begin{align}
\langle \muh_6^3 \rangle_{\text{B/2,N}} &=\frac1{\Dh\Dh_1^2}\Big[(\bh_{65}+1)+(\bh_{64}+1) -\kappah_1\kappah_2\left(\prod_{m=1}^5(\bh_{mm}+1)\right)\nonumber\\
&\qquad\qquad \left. -\left(\prod_{m'=1}^3(\bh_{m'm'}+1)\right)\left[ (\bh_{65}+1)^2+(\bh_{65}+1)(\bh_{64}+1)+(\bh_{64}+1)^2 \right]\kappah_1\right]~,
\end{align}
where
\begin{align}
\Dh&=-\left(\prod_{m=1}^5 (\bh_{mm}+1)\right)\left(\prod_{m'=1}^3 (\bh_{m'm'}+1)\right)\kappah_1^2\kappah_2 \nonumber\\
&\quad+\left(\left(\prod_{m'=1}^3 (\bh_{m'm'}+1)\right)(\bh_{64}+1)\kappah_1-1\right)\left(\left(\prod_{m'=1}^3 (\bh_{m'm'}+1)\right)(\bh_{65}+1)\kappah_1-1\right)~,\nonumber\\
\Dh_1&=(\bh_{64}+1)(\bh_{65}+1)-(\bh_{44}+1)(\bh_{55}+1)\kappah_2~,
\end{align}
are two components of the discriminant locus of the B/2 model, and where we have normalized the correlator according to
\eqref{eq:B2natnorm}.
In appendix \ref{app:B2corrslist} we present the full list of the B/2 model correlators, and we verify explicitly that
under the mirror map the correlators are mapped according to
\begin{align}
\llangle H_{m_1}H_{m_2} H_{m_3} \rrangle \quad \leftrightarrow \quad -2(\bh_{65}+1)^4 \la \muh_{m_1}\muh_{m_2}\muh_{m_3} \ra_{\text{B/2,N}}~.
\end{align}
Again, we find complete agreement up to a non-vanishing\footnote{Alternatively, we can think of the relative normalization as an additional component of the
discriminant locus, as the theory develops a singularity at $\bh_{65}=-1$.} relative normalization.

\section{Discussion}
\label{s:discuss}

In this work we have provided evidence in support of the mirror map for deformations of (2,2) theories proposed in \cite{Melnikov:2010sa}.
In particular, we have shown in two key examples that the map exchanges the A/2 model with the B/2 model of the mirror theory
at the level of correlators. Moreover, we found that the equivalence in question is a fairly simple one: the map exchanges the natural observables 
on the two sides of the mirror 
\begin{align}
\label{eq:02mirrcorrs}
H_\rho=\bgamma^\rho \cdot\bsigma \quad \leftrightarrow \quad \muh_\rho = \alphah_\rho Y_0 \mathsf{M}_\rho~, 
\end{align}
where $\bgamma^0\equiv\bdelta$ and $\mathsf{M}$ labels monomials in the equation defining the CY hypersurface.
In particular, \eqref{eq:02mirrcorrs} holds without requiring, for instance, a parameter dependent redefinition. 
Notice that the relative normalization between the two sets of correlators can be absorbed 
through the $\GL(r,\C)$ field redefinitions of the $\sigma$ fields,
that is, by appropriately rescaling $\bgamma^\rho$ and $\bdelta$.
In this final section we employ our results to derive some consequences on the structure of the moduli space for the theories under study.

We start by showing that the (0,2) moduli space does not exhibit the sort of splitting that would generalize the 
structure on the (2,2) locus. 
A counterexample to such splitting is provided, for instance, by the reflexively plain quintic model discussed in section 
\ref{s:quinticorb}. 
Although solving the full B/2-twisted theory for the $M$ model is beyond the purpose of this work, 
we can restrict our attention to the subset of the $J$-deformations defined by
\begin{align}
\label{eq:Borigmol}
J_\rho &=\frac{j_{0\rho}\alpha_\rho}{\alpha_0}(b_{\rho\rho}+1) X_0X_\rho^4 +j_{0\rho}X_0 \prod_{\rho'\neq \rho} X_{\rho'}~,		&\rho,\rho'&=1,\dots,5~,
\end{align}
and evaluate the correlators at the LGO point, where $X_0$ assumes a non-zero vev. 
This choice for the (0,2) superpotential, 
up to relabeling the various quantities entering \eqref{eq:Borigmol}, formally describes the identical expression we found on the mirror side.
Thus, for this example, $b_{\rho\rho}$, $\rho=1,\dots,5$, enter explicitly in the expressions for correlators of both the A/2- and the B/2-twisted theories 
and, even in this relatively simple example, the bundle moduli do not split into A/2 and B/2 model bundle moduli. 
This shows that in general the GLSM moduli space does not exhibit a product structure. 

Another question is whether there are examples where the B/2 model is itself not classical, meaning, 
it admits non-trivial instanton corrections. 
It has been argued \cite{McOrist:2008ji}
that for linear models with a Landau-Ginzburg phase these corrections do not occur, but what about more general models?
The mirror map provides us with a partial answer. 
It is apparent from the Coulomb branch computations we adopted in this work\footnote{The same holds for the techniques along the lines
of \cite{Closset:2015rna}.} that
the A/2 model admits dependence on the K\"ahler parameters $\kappa$ and the bundle moduli $b$,
while no dependence on $\kappah$ is possible.
Thus, according to the mirror map, the mirror 
B/2 model will depend only on the complex structure parameters $\kappah$ and the bundle moduli $b$, and
it cannot admit instanton corrections. Applying the same reasoning to the mirror model, 
we reach the conclusion that the B/2-twisted theory of the original model is classical as well.
Hence, the moduli space of any model to which these techniques apply does in fact exhibit a partial splitting: 
while the bundle moduli generically play a role in both twisted theories, complex structure and K\"ahler parameters remain a prerogative 
of the B/2 and A/2 models, respectively.

This simple argument, although quite powerful, is subject to two caveats.
First, in non-reflexively plain models, we are forced to work on a subfamily of the full moduli space. On the B/2 model side, this is manifest
through the fact that at the LGO point, where we performed such computations, some of the complex structure/bundle moduli are
forced to be frozen, as the corresponding operators appear in
twisted sectors. In order to study a larger subset of the moduli space it appears necessary to 
employ, if available, a different description of the same CFT where at least some of such moduli are not frozen. 
A systematic study along this lines might help unveiling the structure of the non-reflexive subset of the moduli space. 
For instance, techniques to evaluate B and B/2 model correlators in
hybrid models \cite{Bertolini:2013xga,Bertolini:2017lcz} have been recently developed \cite{Bertolini:2018now},
and these could be employed to gain insights into this larger set of theories.
While we expect a dependence on non-diagonal, but linear, $E$-parameters,
non-linear $E$-parameters seem not to affect A/2-twisted $V$ model correlators \cite{Donagi:2011uz,Donagi:2011va,Donagi:2014koa}. 
However, the situation is more subtle for A/2-twisted $M$ models, where the supersymmetry constraint relates $E$ and $J$ parameters.

Second, the mirror map, as currently formulated, comprises only hypersurfaces in toric varieties. 
It is tempting to conjecture that also for more general models a subset of the B/2 moduli space is protected
by worldsheet instanton effects, and it would be desirable to test this prediction.
Finally, it should be possible, through a deeper understanding of
the combinatorics involved, to extend the mirror map to (0,2) deformations of CICY in toric varieties \cite{Batyrev:1994pg,Batyrev:2007cq}.

\appendix

\allowdisplaybreaks

\section{Correlators for the octic model}
\label{app:corrslists}

In this appendix we collect the full list of correlators in both the A/2 model and B/2 mirror model for the example we solved in section \ref{s:octic}.

\subsection{A/2 model}
\label{app:corrslistsA2}

Let us introduce some notation to simplify the appearance of the result. 
We define $\beta_{m\rho}\equiv b_{m\rho}+1$, for $m,\rho\neq0$ and $\beta_{00}\equiv1$, as well as the products
\begin{align}
\prod_{\rho} \beta_{\rho\rho}&\equiv \prod_{\rho=1}^5 \beta_{\rho\rho}~,		&\prod_{\rho'} \beta_{\rho'\rho'}&=\prod_{\rho'=1}^3 \beta_{\rho'\rho'}~.
\end{align}
In terms of these, the principal component of the discriminant locus takes the form
\begin{align}
D&=-\left(\prod_{\rho} \beta_{\rho\rho}\right)\left(\prod_{\rho'} \beta_{\rho'\rho'}\right)\kappa_1^2\kappa_2 
+\left(\left(\prod_{\rho'} \beta_{\rho'\rho'}\right)\beta_{46}\kappa_1-1\right)\left(\left(\prod_{\rho'} \beta_{\rho'\rho'}\right)\beta_{56}\kappa_1-1\right)~,
\end{align}
while the additional component of the discriminant locus instead reads
\begin{align}
D_1&=\beta_{46}\beta_{56}-\beta_{44}\beta_{55}\kappa_2~.
\end{align}
Finally, in the following we will use the indices $i,j,k=0,1,2,3$. With these conventions, we can now present the full list of correlators.
\begin{align}
\llangle H_i H_j H_k \rrangle &=(-1)^{\delta_{0,i}+\delta_{0,j}+\delta_{0,k}} \frac{2\beta_{56}^4}{\beta_{ii}\beta_{jj}\beta_{kk}D}~,\nonumber\\
\llangle H_i H_j H_6 \rrangle &=(-1)^{\delta_{0,i}+\delta_{0,j}} \frac{2\beta_{56}^4\left(\prod_{\rho'} \beta_{\rho'\rho'}\right)\kappa_1}{\beta_{ii}\beta_{jj}D}~,\nonumber\\
\llangle H_i H_6^2 \rrangle &=-(-1)^{\delta_{0,i}} \frac{2\beta_{56}^4\left[ 1-\left(\prod_{\rho'}\beta_{\rho'\rho'}\right)(\beta_{46}+\beta_{56}) \kappa_1 \right]}{\beta_{ii}DD_1}~,\nonumber\\
\llangle H_6^3 \rrangle &=-\frac{2\beta_{56}^4\left[ \beta_{46}+\beta_{56} 
 -\left(\prod_{\rho}\beta_{\rho\rho}\right) \kappa_1\kappa_2 -\left(\prod_{\rho'}\beta_{\rho'\rho'}\right)\left( \beta_{46}^2 +\beta_{46}\beta_{56} + \beta_{56}^2\right) \kappa_1\right]}{DD_1^2}~,\nonumber\\
\llangle H_i H_j H_5 \rrangle &=(-1)^{\delta_{0,i}+\delta_{0,j}} \frac{2\beta_{56}^4 \left(1-\left(\prod_{\rho'}\beta_{\rho'\rho'}\right)\beta_{56}\kappa_1\right)}{\beta_{ii}\beta_{jj}\beta_{55}D}~,\nonumber\\
\llangle H_i H_5^2 \rrangle &=(-1)^{\delta_{0,i}} \frac{2\beta_{56}^4}{\beta_{ii}\beta_{55}^2DD_1} \left[2\left(\prod_{\rho} \beta_{\rho\rho}\right)\beta_{56}\kappa_1\kappa_2\right. \nonumber\\
&\qquad\qquad\qquad \left.-\left(\prod_{\rho'} \beta_{\rho'\rho'}\right)(\beta_{46}-\beta_{56})\beta_{56}^2
\kappa_1-\beta_{44}\beta_{55}\kappa_2+\beta_{56}(\beta_{46}-\beta_{56})\right]~,\nonumber\\
\llangle  H_5^3 \rrangle &=\frac{2\beta_{56}^4}{\beta_{55}^3DD_1^2}\left[ \beta_{44}^2\beta_{55}^2 \kappa_2^2 
  -3\left(\prod_{\rho} \beta_{\rho\rho}\right) \beta_{44}\beta_{55}\beta_{56}\kappa_1\kappa_2^2 \right.\nonumber\\
&\qquad\qquad\quad +\beta_{44}\beta_{55}\beta_{56}(3\beta_{56}-2\beta_{46})\kappa_2 
 -\left(\prod_{\rho} \beta_{\rho\rho}\right)\beta_{56}^2(4\beta_{56}-3\beta_{46})\kappa_1\kappa_2\nonumber\\
&\qquad\qquad\quad \left. -\beta_{56}^2(\beta_{56}-\beta_{46})^2\left(\left(\prod_{\rho'} \beta_{\rho'\rho'}\right)\beta_{56}\kappa_1-1\right) \right]~,\nonumber\\
\llangle H_i H_j H_4 \rrangle &=(-1)^{\delta_{0,i}+\delta_{0,j}}\frac{2\beta_{56}^4\left(1-\left( \prod_{\rho'} \beta_{\rho'\rho'}\right)\beta_{46}\kappa_1 \right)}{\beta_{ii}\beta_{jj}\beta_{44}D}~,\nonumber\\
\llangle H_i H_4^2 \rrangle &=(-1)^{\delta_{0,i}} \frac{2\beta_{56}^4}{\beta_{ii}\beta_{44}^2DD_1} \left[ \beta_{46}\beta_{56}
+2\left(\prod_{\rho=1}^5 \beta_{\rho\rho}\right)\beta_{46}\kappa_1\kappa_2\right.\nonumber\\
&\qquad\qquad\qquad \left.+ \beta_{46}^2 \left( -1+\left(\prod_{\rho'=1}^3\beta_{\rho'\rho'}\right)(\beta_{46}-\beta_{56})\kappa_1 \right)-\beta_{44}\beta_{55}\kappa_2\right]~,\nonumber\\
\llangle H_4^3 \rrangle&= \frac{2\beta_{56}^4}{\beta_{44}^3DD_1^2}\left[ \beta_{46}^2 (\beta_{46}-\beta_{56})^2
-\left(\prod_{\rho'} \beta_{\rho'\rho'}\right)\beta_{46}^3(\beta_{46}-\beta_{56})^2 \kappa_1\right.\nonumber\\
&\qquad\qquad\qquad +\beta_{44}\beta_{55}\beta_{46}(3\beta_{46}-2\beta_{56})\kappa_2+  \left(\prod_{\rho} \beta_{\rho\rho}\right)\beta_{46}^2(-4\beta_{46}+3\beta_{56})\kappa_1\kappa_2\nonumber\\
&\qquad\qquad\qquad\left.+\beta_{44}^2\beta_{55}^2\kappa_2^2-3 \left(\prod_{\rho} \beta_{\rho\rho}\right)\beta_{44}\beta_{55}\beta_{46}\kappa_1\kappa_2^2\right]~,\nonumber\\
\llangle H_4^2H_5 \rrangle&=\frac{2\beta_{56}^4\kappa_2}{\beta_{44}DD_1^2}
\left[ \beta_{46}^2  +\left(\prod_{\rho'} \beta_{\rho'\rho'}\right)\beta_{46}^3\kappa_1 -\left(\prod_{\rho} \beta_{\rho\rho}\right)(2\beta_{46}+\beta_{56})\kappa_1\kappa_2+\beta_{44}\beta_{55}\kappa_2\right]~,\nonumber\\
\llangle H_4H_5^2 \rrangle&=\frac{2\beta_{56}^4\kappa_2}{\beta_{55}DD_1^2}
\left[ \beta_{56}^2 +\left(\prod_{\rho} \beta_{\rho'\rho'}\right)\beta_{56}^3\kappa_1 -\left(\prod_{\rho} \beta_{\rho\rho}\right)(2\beta_{56}+\beta_{46})\kappa_1\kappa_2
+\beta_{44}\beta_{55}\kappa_2\right]~,\nonumber\\
\llangle H_iH_4H_5 \rrangle&=(-1)^{\delta_{i,0}} \frac{2\beta_{56}^4\kappa_2\left(-1+\left(\prod_{\rho'} \beta_{\rho'\rho'}\right)(\beta_{46}+\beta_{56})\kappa_1\right)}{\beta_{ii}DD_1}~,\nonumber\\
\llangle H_4H_5H_6 \rrangle&=-\frac{2\beta_{56}^4\kappa_2}{DD_1^2}\left[ \beta_{46}+\beta_{56} 
-\left(\prod_{\rho}\beta_{\rho\rho}\right) \kappa_1\kappa_2 -\left(\prod_{\rho'}\beta_{\rho'\rho'}\right)\left( \beta_{46}^2 +\beta_{46}\beta_{56} + \beta_{56}^2\right) \kappa_1\right]~,\nonumber\\
\llangle H_4^2H_6 \rrangle&=\frac{2\beta_{56}^4}{\beta_{44}^2 DD_1^2}
\left[-\beta_{46}^2(\beta_{46}-\beta_{56})  
+ \left(\prod_{\rho'} \beta_{\rho'\rho'}\right)\beta_{46}^3(\beta_{46}-\beta_{56})\kappa_1 -2\beta_{44}\beta_{55}\beta_{46}\kappa_2 \right.\nonumber\\
&\qquad\qquad\qquad \left. +3\left(\prod_{\rho} \beta_{\rho\rho}\right)\beta_{46}^2\kappa_1\kappa_2
+\left(\prod_{\rho} \beta_{\rho\rho}\right)\beta_{44}\beta_{55}\kappa_1\kappa_2^2 \right]~,\nonumber\\
\llangle H_4H_6^2 \rrangle&=-\frac{2\beta_{56}^4}{\beta_{44}DD_1^2}
\left[ -\beta_{46}^2 
+\left(\prod_{\rho'} \beta_{\rho'\rho'}\right)\beta_{46}^3\kappa_1 +\left(\prod_{\rho} \beta_{\rho\rho}\right)(2\beta_{46}+\beta_{56})\kappa_1\kappa_2
-\beta_{44}\beta_{55}\kappa_2\right]~,\nonumber\\
\llangle H_5^2H_6 \rrangle&=\frac{2\beta_{56}^4}{\beta_{55}^2 DD_1^2}
\left[-\beta_{56}^2(\beta_{56}-\beta_{46})
+\left(\prod_{\rho'=1}^3 \beta_{\rho'\rho'}\right)\beta_{56}^3(\beta_{56}-\beta_{46})\kappa_1 -2\beta_{44}\beta_{55}\beta_{56}\kappa_2 \right.\nonumber\\
&\qquad\qquad\qquad \left. +3\left(\prod_{\rho} \beta_{\rho\rho}\right)\beta_{56}^2\kappa_1\kappa_2
+\left(\prod_{\rho=1}^5 \beta_{\rho\rho}\right)\beta_{44}\beta_{55}\kappa_1\kappa_2^2 \right]~,\nonumber\\
\llangle H_5H_6^2 \rrangle&=-\frac{2\beta_{56}^4}{\beta_{55}DD_1^2}
\left[ -\beta_{56}^2 +\left(\prod_{\rho'} \beta_{\rho'\rho'}\right)\beta_{56}^3\kappa_1 +\left(\prod_{\rho} \beta_{\rho\rho}\right)(2\beta_{56}+\beta_{46})\kappa_1\kappa_2
-\beta_{44}\beta_{55}\kappa_2\right]~,\nonumber\\
\llangle H_i H_4 H_6 \rrangle&=-(-1)^{\delta_{i,0}}  \frac{2\beta_{56}^4\left[ -\beta_{46}+\left(\prod_{\rho'} \beta_{\rho'\rho'}\right)\beta_{46}^2\kappa_1  
+\left(\prod_{\rho} \beta_{\rho\rho}\right)\kappa_1\kappa_2\right]}{\beta_{ii}\beta_{44} D D_1}~,\nonumber\\
\llangle H_i H_5 H_6 \rrangle&=-(-1)^{\delta_{i,0}}  \frac{2\beta_{56}^4\left[ -\beta_{56}+\left(\prod_{\rho'} \beta_{\rho'\rho'}\right)\beta_{56}^2\kappa_1  
+\left(\prod_{\rho} \beta_{\rho\rho}\right)\kappa_1\kappa_2\right]}{\beta_{ii}\beta_{55} D D_1}~.
\end{align}

\subsection{B/2 model}
\label{app:B2corrslist}

We introduce here as well a similar notation as we did in the previous section. 
We define the parameters $\betah_{\rho m }\equiv \bh_{\rho m}+1$, for $\rho,m\neq0$, with $\betah_{00}\equiv1$, 
as well as 
\begin{align}
\prod_{m} \betah_{mm}&\equiv \prod_{m=1}^5 \betah_{mm}~,		&\prod_{m'} \betah_{m'm'}&=\prod_{m'=1}^3 \betah_{m'm'}~.
\end{align}
The principal component of the discriminant locus reads
\begin{align}
\Dh&=-\left(\prod_{m} \betah_{mm}\right)\left(\prod_{m'} \betah_{m'm'}\right)\kappah_1^2\kappah_2 
+\left(\left(\prod_{m'} \betah_{m'm'}\right)\betah_{64}\kappah_1-1\right)\left(\left(\prod_{m'} \betah_{m'm'}\right)\betah_{65}\kappah_1-1\right)~,
\end{align}
while we indicate as
\begin{align}
\Dh_1=\betah_{64}\betah_{65}-\betah_{44}\betah_{55}\kh_2
\end{align} 
the additional component of the discriminant locus.
We present below the full list of correlators, where $i,j,k=0,\dots,3$.
\begin{align}
\langle \muh_i\muh_j\muh_k \rangle &=-(-1)^{\delta_{i,0}+\delta_{j,0}+\delta_{k,0}}\frac{1}{\betah_{ii}\betah_{jj}\betah_{kk}\Dh}~,\nonumber\\
\langle \muh_i\muh_j \muh_6 \rangle &= - (-1)^{\delta_{i,0}+\delta_{j,0}}\frac{\kappah_1 \prod_{m'}\betah_{m'm'}}{\betah_{ii}\betah_{jj}\Dh} ~,\nonumber\\
\langle \muh_i \muh_6^2 \rangle &=-(-1)^{\delta_{i,0}} \frac{\left(\prod_{m'}\betah_{m'm'} \right)(\betah_{64}+\betah_{65})\kappah_1-1}{\betah_{ii}\Dh\Dh_1}~,\nonumber\\
\langle \muh_6^3 \rangle &=\frac{\betah_{65}+\betah_{64} -\left(\prod_{m'}\bh_{m'm'}\right)\left[ \betah_{65}^2+\betah_{65}\betah_{64}+\betah_{64}^2 \right]\kappah_1-\left(\prod_m \betah_{mm}\right)\kappah_1\kappah_2}{\Dh\Dh_1^2}~,\nonumber\\
\langle \muh_i\muh_j\muh_5 \rangle&= (-1)^{\delta_{i,0}+\delta_{j,0}}\frac{-1+\left(\prod_{m'}\betah_{m'm'}\right)\betah_{65}\kappah_1}{\betah_{ii}\betah_{jj}\betah_{55}\Dh}~,\nonumber\\
\langle \muh_i\muh_5^2 \rangle&=-(-1)^{\delta_{i,0}}\frac1{\betah_{ii}\betah_{55}^2\Dh\Dh_1}\left[(\betah_{64}-\betah_{65})\betah_{65} 
- \left(\prod_{m'}\betah_{m'm'}\right)\betah_{65}^2(\betah_{64}-\betah_{65})\kappah_1\right.\nonumber\\
&\qquad\qquad\qquad\qquad\qquad\quad \left. -\betah_{44}\betah_{55}\kappah_2+2 \left(\prod_{m}\betah_{mm}\right)\betah_{65}\kappah_1\kappah_2 \right]~,\nonumber\\
\langle \muh_5^3 \rangle&=-\frac1{\betah_{55}^3\Dh\Dh_1^2}\left[ \betah_{65}^2(\betah_{64}-\betah_{65})^2
-\left(\prod_{m'}\betah_{m'm'}\right)\betah_{65}^3(\betah_{64}-\betah_{65})^2\kappah_1  \right.\nonumber\\
&\qquad\qquad\qquad -\betah_{44}\betah_{55}\betah_{65}(2\betah_{64}-3\betah_{65}) \kappah_2 
+\left(\prod_{m}\betah_{mm}\right)\betah_{65}^2(3\betah_{64}-4\betah_{65})\kappah_1\kappah_2 \nonumber\\
&\qquad\qquad\qquad \left.+\betah_{55}^2\betah_{44}^2\kappah_2^2 -3\left(\prod_{m}\betah_{mm}\right)\betah_{44}\betah_{55}\betah_{65}\kappah_1\kappah_2^2 \right]~,\nonumber\\
\langle \muh_i\muh_j\muh_4 \rangle&=(-1)^{\delta_{i,0}+\delta_{j,0}} \frac{-1+\left(\prod_{m'}\betah_{m'm'}\right)\betah_{64}\kappah_1}{\betah_{ii}\betah_{jj}\betah_{44}\Dh}~,\nonumber\\
\langle \muh_i\muh_4^2 \rangle&=-(-1)^{\delta_{i,0}} \frac1{\betah_{ii}\betah_{44}^2\Dh\Dh_1}\left[(\betah_{65}-\betah_{64})\betah_{64} 
- \left(\prod_{m'}\betah_{m'm'}\right)\betah_{64}^2(\betah_{65}-\betah_{64})\kappah_1\right.\nonumber\\
&\qquad\qquad\qquad\qquad\qquad\quad \left. -\betah_{44}\betah_{55}\kappah_2+2 \left(\prod_{m}\betah_{mm}\right)\betah_{64}\kappah_1\kappah_2 \right]~,\nonumber\\
\langle \muh_4^3 \rangle&=-\frac1{\betah_{44}^3\Dh\Dh_1^2}\left[ \betah_{65}^2(\betah_{65}-\betah_{64})^2
-\left(\prod_{m'}\betah_{m'm'}\right)\betah_{64}^3(\betah_{65}-\betah_{64})^2\kappah_1  \right.\nonumber\\
&\qquad\qquad\qquad -\betah_{44}\betah_{55}\betah_{64}(2\betah_{65}-3\betah_{64}) \kappah_2 
+\left(\prod_{m}\betah_{mm}\right)\betah_{64}^2(3\betah_{65}-4\betah_{64})\kappah_1\kappah_2 \nonumber\\
&\qquad\qquad\qquad \left.+\betah_{55}^2\betah_{44}^2\kappah_2^2 -3\left(\prod_{m}\betah_{mm}\right)\betah_{44}\betah_{55}\betah_{64}\kappah_1\kappah_2^2 \right]~,\nonumber\\
\langle \muh_4^2\muh_5 \rangle &= -\frac{\kappah_2}{\betah_{44}\Dh\Dh_1^2} \left[ \betah_{64}^2 +\betah_{44}\betah_{55}\kappah_2  
 -\left(\prod_{m'}\betah_{m'm'}\right) \betah_{64}^3\kappah_1 -\left(\prod_{m}\betah_{mm}\right)(2\betah_{64}+\betah_{65})\kappah_1\kappah_2 \right]~,\nonumber\\
\langle \muh_4\muh_5^2 \rangle &= -\frac{\kappah_2}{\betah_{55}\Dh\Dh_1^2} \left[ \betah_{65}^2 +\betah_{44}\betah_{55}\kappah_2 
-\left(\prod_{m'}\betah_{m'm'}\right) \betah_{65}^3\kappah_1 -\left(\prod_{m}\betah_{mm}\right)(2\betah_{65}+\betah_{64})\kappah_1\kappah_2 \right]~,\nonumber\\
\langle \muh_i\muh_4\muh_5 \rangle &=-(-1)^{\delta_{i,0}}\frac{-\kappah_2+\left(\prod_{m'}\betah_{m'm'}(\betah_{64}+\betah_{65})\kappah_1\kappah_2\right)}{\betah_{ii}\Dh\Dh_1}~,\nonumber\\
\langle \muh_4\muh_5\muh_6 \rangle &=\frac{\kappah_2\left[\betah_{65}+\betah_{64} -\left(\prod_{m'}\betah_{m'm'}\right)\left( \betah_{65}^2+\betah_{65}\betah_{64}+\betah_{64}^2 \right)\kappah_1-\left(\prod_m \betah_{mm}\right)\kappah_1\kappah_2 \right]}{\Dh\Dh_1^2}~,\nonumber\\
\langle \muh_4^2\muh_6 \rangle &= -\frac1{\betah_{44}^2\Dh\Dh_1^2} \left[ \betah_{64}^2(\betah_{65}-\betah_{64}) 
-2\betah_{44}\betah_{55}\betah_{64}\kappah_2 +\left(\prod_{m'}\betah_{m'm'}\right) \betah_{64}^3(\betah_{64}-\betah_{65})\kappah_1 \right.\nonumber\\
&\qquad\qquad\qquad \left. +3\left(\prod_{m}\betah_{mm}\right)\betah_{64}^2\kappah_1\kappah_2
+\left(\prod_{m}\betah_{mm}\right)\betah_{44}\betah_{55}\kappah_1\kappah_2^2    \right]~,\nonumber\\
\langle \muh_4\muh_6^2 \rangle &= -\frac1{\betah_{44}\Dh\Dh_1^2} \left[ \betah_{64}^2 +\betah_{44}\betah_{55}\kappah_2 
 -\left(\prod_{m'}\betah_{m'm'}\right) \betah_{64}^3\kappah_1 -\left(\prod_{m}\betah_{mm}\right)(2\betah_{64}+\betah_{65})\kappah_1\kappah_2 \right]~,\nonumber\\
 \langle \muh_5^2\muh_6 \rangle &= -\frac1{\betah_{55}^2\Dh\Dh_1^2} \left[ \betah_{65}^2(\betah_{64}-\betah_{65}) 
-2\betah_{44}\betah_{55}\betah_{65}\kappah_2 +\left(\prod_{m'}\betah_{m'm'}\right) \betah_{65}^3(\betah_{65}-\betah_{64})\kappah_1  \right.\nonumber\\
&\qquad\qquad\qquad \left. +3\left(\prod_{m}\betah_{mm}\right)\betah_{65}^2\kappah_1\kappah_2
+\left(\prod_{m}\betah_{mm}\right)\betah_{44}\betah_{55}\kappah_1\kappah_2^2    \right]~,\nonumber\\
\langle \muh_5\muh_6^2 \rangle &= -\frac1{\betah_{55}\Dh\Dh_1^2} \left[ \betah_{65}^2 +\betah_{44}\betah_{55}\kappah_2 
 -\left(\prod_{m'}\betah_{m'm'}\right) \betah_{65}^3\kappah_1 -\left(\prod_{m}\betah_{mm}\right)(2\betah_{65}+\betah_{64})\kappah_1\kappah_2 \right]~,\nonumber\\
\langle \muh_i\muh_4\muh_6 \rangle&=-(-1)^{\delta_{i,0}} \frac{\betah_{64}-\left(\prod_{m'}\betah_{m'm'}\right)\betah_{64}^2\kappah_1
-\left(\prod_{m}\betah_{mm}\right)\kappah_1\kappah_2}{\betah_{ii}\betah_{44}\Dh\Dh_1}~,\nonumber\\
\langle \muh_i\muh_5\muh_6 \rangle&= -(-1)^{\delta_{i,0}}\frac{\betah_{65}-\left(\prod_{m'}\betah_{m'm'}\right)\betah_{65}^2\kappah_1
-\left(\prod_{m}\betah_{mm}\right)\kappah_1\kappah_2}{\betah_{ii}\betah_{55}\Dh\Dh_1}~.
\end{align}

\bibliographystyle{./utphys}
\bibliography{./bigref}

\end{document}